\documentclass[preprint2]{proto}
\usepackage{times}
\usepackage{xcolor}
\usepackage{graphicx}
\usepackage{float}
\usepackage{dblfloatfix}
\usepackage{amsmath}
\usepackage{enumitem}
\usepackage{hyperref}
\usepackage{longtable}
\usepackage{soul}

\graphicspath{{./}{Figures/}}

\begin{document}

\title{\textbf{\LARGE THE TRANSITION FROM THE KUIPER BELT TO THE JUPITER-FAMILY COMETS}}

\author {\textbf{\large Wesley C. Fraser}}
\affil{\small\em Herzberg Astronomy and Astrophysics Research Centre, National Research Council, \\ 5071 West Saanich Road, Victoria, BC, V8T 1E7, Canada}
\affil{\small\em Department of Physics and Astronomy, University of Victoria, Victoria, BC, V8P 5C2, Canada}

\author {\textbf{\large Luke Dones}}
\affil{\small\em Southwest Research Institute, 1050 Walnut St., Suite 300, Boulder, CO, 80504, USA}

\author {\textbf{\large Kathryn Volk}}
\affil{\small\em Lunar and Planetary Laboratory, The University of Arizona, 1629 E University Blvd, Tucson, AZ, 85721}
\affil{\small\em Planetary Science Institute, 1700 East Fort Lowell Rd., Suite 106, Tucson, Arizona, 85719}

\author {\textbf{\large Maria Womack}}
\affil{\small\em National Science Foundation, 2415 Eisenhower Ave, Alexandria, VA 22314 \\ Department of Physics, University of Central Florida, Orlando, FL 32816}

\author {\textbf{\large David Nesvorn\'y}}
\affil{\small\em Southwest Research Institute, 1050 Walnut St., Suite 300, Boulder, CO, 80504, USA}

\begin{abstract}

\begin{list}{ } {\rightmargin 1in}
\baselineskip = 11pt
\parindent=1pc
{\small 
Kuiper Belt Objects, or more generally Trans-Neptunian Objects (TNOs), are planetesimals found beyond the orbit of Neptune. Some TNOs evolve onto Neptune-crossing orbits and become Centaurs. Many Centaurs, in turn, reach Jupiter-crossing orbits and become Jupiter-family comets (JFCs). TNOs are the main source of the JFCs. TNOs offer a different window than the JFCs, of more primordial bodies and over a different size and temperature range. It is in that context that this chapter is written. Here we discuss the dynamical pathways taken from the trans-Neptunian region to the JFCs, and the most important properties of TNOs that relate to the JFC population, including considerations of their origins, compositions, morphologies, and size distributions. We relate these properties to the JFCs whenever possible. We reflect on a few key outstanding issues regarding our incomplete knowledge of TNOs as they pertain to the Centaurs and JFC populations. We finish with a short discussion of notable new and upcoming facilities and the impacts they will have regarding these outstanding questions.
\\~\\~\\~}
\end{list}
\end{abstract}  

\section{\textbf{Introduction}}
\label{sec:intro}

\citet{Oort1950} proposed the first, and for some time the only, reservoir for comets. He noted that, of nineteen long-period comets with well-determined orbits and semi-major axes $a \gtrsim 1,000$~au, ten had $a > 20,000$~au before they entered the planetary region. These ``near-parabolic'' comets had a wide range of inclinations to the ecliptic, both prograde and retrograde. Oort proposed that the solar system was surrounded by a roughly spherical cloud of $\approx 10^{11}$ comets that extended beyond $\approx 10^5$~au from the Sun, about half the distance to the nearest star.

In the wonderful words of \citet{Kazimirchak-Polonskaya1972}, the major planets can indeed serve as ``powerful transformers of cometary orbits,'' with some evolving from near-parabolic orbits onto orbits like those of Jupiter-family Comets (JFCs), which we discuss in Section~\ref{sec:eclipticcomets} \citep{Everhart1972}. However, the ``capture'' of Oort Cloud comets into short-period orbits is far too inefficient to produce the observed number of JFCs \citep{Joss1973}. This mismatch suggested that there must be another source of short-period comets. Several astronomers, notably \citet{Edgeworth1949} and \citet{Kuiper51a}, suggested the existence of a belt of low-inclination comet-like bodies beyond Neptune's orbit, $\approx$ 35--50~au from the Sun (see \citet{Davies2008} and \citet{Fernandez2020} for historical accounts of such ideas, which percolated even before the discovery of Pluto in 1930). \citet{Fernandez1980} calculated that such a belt could be the source of short-period comets if it contained bodies up to the mass of Ceres ($\approx 10^{24}$~g, or 7\% the mass of Pluto). \citet{Duncan1988} performed extensive numerical integrations and confirmed that only a low-inclination source could produce enough JFCs and proposed that this source be called the Kuiper Belt. The first Kuiper Belt Object, 15760 Albion (1992~QB$_1$), was discovered four years later \citep{Jewitt1993}.\footnote{The community has referred to the objects beyond Neptune as Kuiper Belt Objects and Trans-Neptunian Objects, largely interchangeably. We adopt the latter here as a more generalized and less historically charged term.}  \citet{Duncan1988} also remarked that Chiron, the only Centaur known at the time, ``may well be a bright member of the parent population of [short-period] comets.''
Hundreds of Centaurs are now known, including P/2019 LD$_2$ (ATLAS), an active body which passed within $\approx 0.1$~au of Jupiter in 2017 and will do again in 2028 and 2063 \citep{Steckloff2020}. The 2063 encounter, the closest of the three, is expected to place the comet on an orbit with a semimajor axis $a \approx 4$~au and a perihelion distance $q  < 3$~au, making it a Jupiter-family comet.

Because of the inverse-fourth-power dependence on reflected flux density with distance, the known Trans-Neptunian Objects (TNOs) are typically much larger than the Centaurs, which, in turn, are generally larger than the nuclei of JFCs. Most known TNOs have diameters of order 100 km, ranging up to Pluto and Eris (2300--2400~km); Centaurs are generally in the range of 10 to 100~km, with Chariklo the largest at 300~km; the nuclei of Jupiter-family comets range from $\approx 0.3$~km up to more than 10~km. These size differences should be kept in mind when comparing different classes of bodies. 

In the remainder of this section, we summarize basic orbital properties of TNOs, Centaurs, and comets, particularly the JFCs. Section 2 discusses how the Kuiper Belt attained its complex structure and the orbital evolution of TNOs to Centaurs and JFCs. In Section 3, we discuss the compositions of TNOs and Centaurs revealed by broadband colors and reflectance spectroscopy. Section 4 treats the active Centaurs, whose activity must be driven by substances more volatile than water ice; the issue of whether the frequent appearance of bilobed structures in JFCs reflects morphological evolution after leaving the Kuiper Belt or dates to the formation of these bodies billions of years ago; and the size distributions of TNOs, Centaurs, and JFCs. Section 5 concludes with a discussion of outstanding questions and future observations that may clarify many of the issues we discuss. This chapter addresses a variety of different topics regarding TNOs. Rather than providing a detailed review of a specific topic, this chapter aims to provide an overview of the most relevant TNO science relating to the field of cometary astronomy.

Many chapters in {\em Comets III} discuss subjects which we have no room to describe. We particularly note the chapters on the dynamical populations of comet reservoirs by Kaib and Volk; planetesimal and comet formation by Simon et al.; the chapters on cometary nuclei by Guilbert-Lepoutre et al., Pajola et al., Filacchione et al., and Knight et al.; past and future comet missions by Snodgrass et al.; and the chapter by Jewitt \& Hsieh on transition objects, especially their discussion of asteroids on cometary orbits.

\subsection{Meet the Kuiper Belt}
\label{sec:kuiperpops}

\begin{figure}[!htp]
\begin{center}
\includegraphics[width=0.49\textwidth]{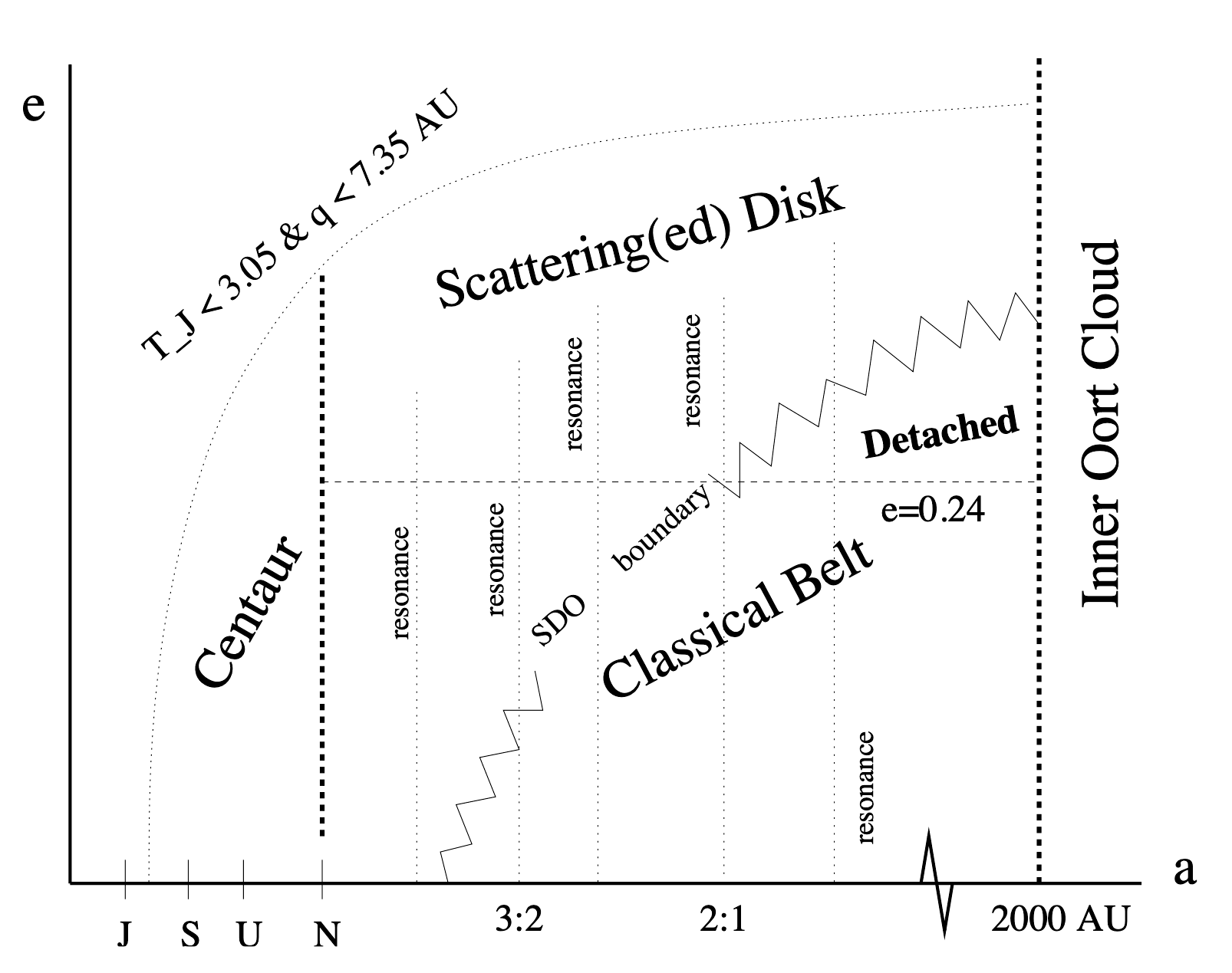}
\caption{\textcolor{black}{Nomenclature scheme for objects with semi-major axes beyond Jupiter's orbit \citep[][from]{Gladman2008}. This taxonomy primarily focuses on TNOs, and, by design, excludes most JFCs. We use a similar classification.}} 
\label{fig:gladman-taxonomy}
\end{center}
\end{figure}

\begin{figure}[!ht]
\centering
\includegraphics[width=0.49\textwidth]{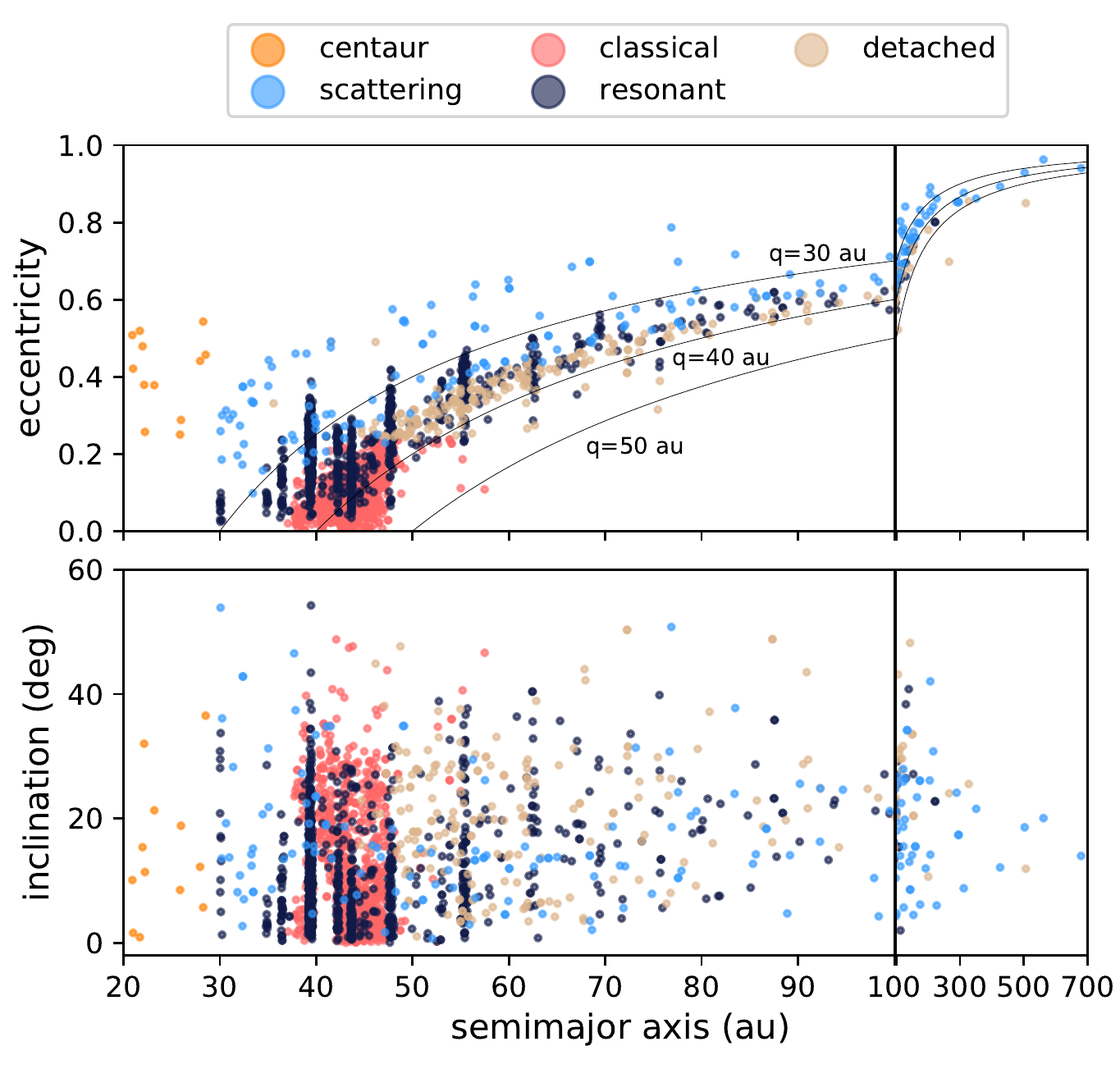}
\caption{\textcolor{black}{Eccentricity and inclination vs.\ semimajor axis of $\sim1800$ observed TNOs color-coded by dynamical class; the set of TNOs is the full OSSOS sample \citep{Bannister2018} combined with a subset of other TNOs in the Minor Planet Center database with classifications taken from \cite{Smullen2020}. 
The three solid lines in the top panel labeled with values between 30, 40, and 50 represent lines of constant perihelion distance. Modified from \cite{Gladman2021}. See Figure~8 of \citet{Bernardinelli2022} for a similar plot of discoveries by the Dark Energy Survey.}} 
\label{fig:aei}
\end{figure}

Three decades after the discovery of the first Trans-Neptunian Object (besides Pluto/Charon), the dynamical structure of the Kuiper Belt has proven to be more complex than that of the asteroid belt (see \citealt{Gladman2021} for a recent review). 
TNOs are typically classified based on 10-Myr orbital integrations, as this timescale reveals the population's relevant key dynamical behaviors (a scheme that began with early observational surveys; e.g., \citealt{Elliot2005}).
\textcolor{black}{
\cite{Gladman2008} established a widely-used classification scheme (Figure~\ref{fig:gladman-taxonomy}) that divides outer solar system objects into the following categories:
\begin{itemize}
    \item Jupiter-coupled: Objects with semimajor axes exterior to Jupiter, but with Tisserand parameters with respect to Jupiter\footnote{The Tisserand parameter $T$ of a comet with respect to a planet with semimajor axis $a_P$ is $T = a_P/a + 2 \cos i \sqrt{a (1-e^2)/a_P}$. Here $a$, $e$, and $i$ are the semimajor axis, eccentricity, and inclination of the comet with respect to the planet's orbital plane. The Tisserand parameter is an approximation to the Jacobi integral, which is a constant of motion for the circular restricted three-body problem. For Jupiter-coupled comets, the Tisserand parameter varies more gradually than other orbital parameters, such as the comet's period of revolution around the Sun \citep{Kresak1972, Carusi1987a, Levison1996}.} less than 3.05, consistent with cometary orbits
    \item Centaurs: Non-Trojan objects with semimajor axes between those of Jupiter (5.2~au) and Neptune (30.1~au)
\item Resonant: Objects in mean-motion resonances (MMRs) with Neptune
\item Scattering: Objects whose semimajor axes vary by at least 1.5~au over 10 Myr, which typically coincides with perihelion distances $\lesssim$ 38-40~au
\item Classical: Non-resonant, non-scattering objects with eccentricities $<0.24$, an arbitrary cutoff that restricts this population largely to semimajor axes between 30 and $\approx 50$~au
\item Detached: Objects on stable orbits that fall into none of the previous categories 
\end{itemize}
}
The population of classical objects residing between Neptune's 3:2 and 2:1 resonances (39.4~$<a<$~47.8~au) is often referred to as the `main belt', with classical objects interior and exterior to those resonances referred to as inner and outer classical belt objects.
\textcolor{black}{Figure~\ref{fig:aei} shows a set of classified observed outer solar system objects} that illustrates a few key dynamical features of this population: the prevalence of resonant objects, a bimodal inclination distribution extending to quite high $i$, and a mix of scattering and detached objects with a wide range of perihelion distances in the high-$a$ population.

Populations of dynamically stable objects in Neptune's MMRs were apparent from early surveys \citep[e.g.][]{Chiang2002,Elliot2005} and predicted by early theoretical studies \citep[e.g.][]{Duncan1995,Malhotra1995}. 
Although the 3:2 MMR, which hosts Pluto and numerous ``Plutinos,'' contains more {\em known} TNOs than other resonances, the raw numbers are misleading. This is because bodies in the 3:2 have shorter orbital periods and smaller perihelion distances 
than most TNOs, and so are easier to discover.
Analyses accounting for observational biases imply that a wide range of Neptune's resonances in the classical belt and in more distant regions have large populations, comparable to that of the 3:2 MMR
\citep[e.g.][]{Gladman2012,Adams2014,Volk2016,
Chen2019,Crompvoets2022}. This has important implications for the dynamical history of the outer solar system (see Section~\ref{sec:origins_tnos}).
\textcolor{black}{By the late 1990s, it was clear that the orbits of observed TNOs extended to higher inclinations than did orbits of main-belt asteroids. However, there is a concentration of low-inclination orbits in the classical belt from $a=42-45$~au. }
The bimodal nature of the inclination distribution in the classical Kuiper belt region was first demonstrated by \cite{Brown2001}. 
The low-inclination classical belt objects also tend to have low eccentricities; this sparked the division of TNO populations into categories of dynamically `hot' and `cold' subpopulations [meant to invoke thermodynamics because the `hot' population have larger radial and vertical velocities than 'cold' objects] \citep{Levison2001}. 
The cold population is essentially entirely constrained to $a=42-45$~au (though with possible extensions just beyond 50~au; see, e.g., \citealt{Bannister2018}), while the scattering, detached, and the majority of the resonant populations all fall into the dynamically hot category (the implications of which are discussed in Section~\ref{sec:origins_tnos}).

At large semimajor axes, the observed population is a mix of resonant, scattering, and detached objects spanning an unexpectedly wide range of perihelion distances from 30 to 80~au. 
The existence of a scattered disk \citep{Luu1997, Duncan1997, Volk2008} or scattering population \citep{Gladman2008, Gladman2021} 
with perihelion distances less than $\approx37-40$~au is a natural consequence of objects being fed onto Neptune-crossing orbits, either today or in the early solar system (e.g., \citealt{Levison1997, Lykawka2007}; also see the discussion in the Kaib \& Volk chapter). 
Objects with these smaller perihelion distances can experience perturbations from Neptune at perihelion that are strong enough to cause significant changes in orbital energy, resulting in a random walk in semimajor axis. This creates the fan-like structure in $a-e$ space seen in Figure~\ref{fig:aei}, where objects are dispersed in semimajor axis along lines of constant perihelion distance (the dashed curves). 
Interactions with Neptune's resonances can cause some objects' perihelia to be raised and lowered, causing them to cycle between scattering, resonant, and detached states;
other objects appear to be permanently stranded in high-perihelion detached orbits, some likely via resonant interactions during planet migration (reviewed in, e.g., \citealt{Gomes2008}).
\textcolor{black}{There is a subset of detached objects with such large perihelia (90377 Sedna with $q = 76$~au, 2012 VP$_{113}$ with $q=80$~au; \citealt{Brown2004,Trujillo2014,Sheppard2019}) that they require another explanation beyond the perturbations of the giant planets (see, e.g., \citealt{Gladman2021}). Additional large-$a$, large-$q$ objects that might bridge gaps between the scattering, detached, and perhaps the inner Oort cloud populations are also beginning to be discovered (e.g., 541132 Leleakuhonua (2015 TG$_{387}$) with $q = 65$~au and $a \approx 1000$~au).}
As these `extreme' objects do not contribute to the supply of short-period comets, we will not discuss them at length here (their implications for the solar system's history are touched on in Section~\ref{sec:origins_tnos}, and their possible relationship to the inner Oort cloud is discussed in the Kaib \& Volk chapter).

\subsection{Meet the Comets}
\label{sec:eclipticcomets}

\textcolor{black}{
Most comets likely formed beyond the water ``snow line'' $\approx 3$~au from the Sun [however, the snow line's location changes as the protoplanetary disk evolves \citep{Oberg2011, Marboeuf2014, Harsono2015, Cieza2016, Drozdovskaya2016}] and within $\approx 30$~au \citep{Nesvorny2018} 
in the solar system's initial planetesimal disk. 
}

\textcolor{black}{
Comets are commonly divided into groups based on their orbital periods. In the traditional taxonomy, long-period comets (LPCs) are defined as those with orbital periods $P >$ 200 years, corresponding to $a > 34.2$~au, and short-period, or periodic, comets have $P <$ 200 years. 
The 200-year boundary has no fundamental significance dynamically or compositionally \citep[see, e.g.][]{AHearn1995, Mumma2011}.
}

\textcolor{black}{
Long-period comets are thought to have been gravitationally scattered by interactions with the giant planets from their closer-in formation regions ($\approx 3$--30 au) out to distances between 2,000 and $\sim$~100,000 au to form a collective known as the Oort Cloud \citep[see, e.g.][]{Duncan1987, Vokrouhlicky2019}. At those distances, perturbations external to the solar system, including passing stars and galactic tides, raise and lower comets' perihelion distances in and out of the giant planet region and cause their orbits to take on nearly random orientations (see for example, \citealt{Weissman1996, Dones2004}; also see the chapter by Kaib \& Volk in this volume for further discussion of long-period comets and Oort Cloud formation).
}

\begin{figure}[htbp]
 \centering
\includegraphics[width=0.49\textwidth]{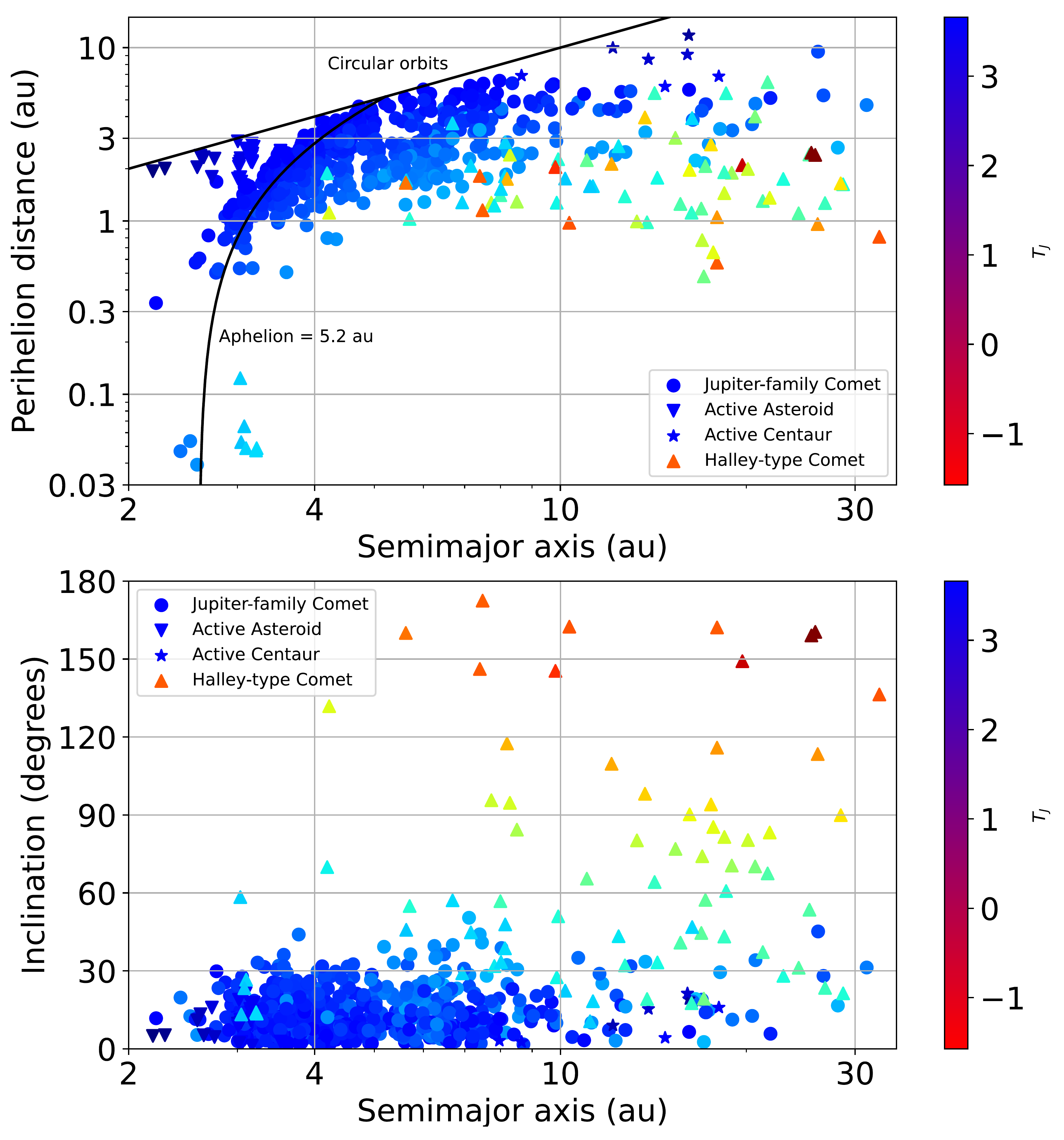}
\caption{\textcolor{black}{Plot of short-period comets (orbital period 
$< 200$~years, equivalent to $a \lesssim 34.2$~au) listed in the JPL Small-Body Database browser. 
We use the definition of \citet{Gladman2008} for Jupiter-family comets (JFCs), Tisserand parameter relative to Jupiter $2 < T_J < 3.05$. JFCs are denoted by circles; active asteroids ($T_J \geq 3.05$, $a<a_J$) by downward-pointing arrows); active Centaurs ($T_J \geq 3.05$, $a \geq a_J$) by stars; and Halley-type comets ($T_J \leq 2$) by upward-pointing arrows. The top panel shows perihelion distance vs.\ semimajor axis, while the bottom panel shows inclination vs.\ semimajor axis.
The color scale shows $T_J$ for each comet (right colorbars). The lines in the top panel denote circular orbits and aphelion distance equal to Jupiter's semimajor axis of 5.2~au. Objects between the lines have orbits entirely within Jupiter's orbit.}}
\label{fig:comets}
\end{figure}

\textcolor{black}{
The short-period comets can be further divided into sub-populations \citep{Horner2003, Seligman2021}. 
Most are classified as Jupiter-family Comets (JFCs), many of which have aphelion distances near Jupiter's distance from the Sun (Figure~\ref{fig:comets}). Most JFCs and Centaurs originate from the dynamically hot trans-Neptunian populations  (Section~\ref{sec:kuiperpops}) and orbit the Sun on prograde orbits with inclinations smaller than $\sim30^\circ$, which is why they are also sometimes referred to as ecliptic comets, e.g., \citealt{Levison1997}. While most JFCs originate from the TNO populations, other populations likely contribute as well; Section~\ref{sec:origins_tnos} discusses this in further detail.
The smaller semimajor axes and inclinations of JFC orbits, which are not consistent with an Oort Cloud origin, pointed the way to the existence of the Kuiper belt/TNO populations prior to their discovery \citep[e.g.][]{Edgeworth1949, Fernandez1980, Duncan1988}. This connection was later confirmed by models of the newly observed reservoir \citep[e.g.][]{Levison1997, Duncan1997}.
}

\textcolor{black}{
Historically, JFCs were often defined as comets with $P < 20$~years ($a < 7.4$~au), although a variety of definitions appear in the literature. JFCs are now often taken to be comets with  Tisserand parameters with respect to Jupiter, $T_J$, between 2 and 3 \citep{Carusi1987a}. This is the definition suggested by \citealt{Levison1997} as being the most indicative of an origin in the Kuiper belt.
Slight variations on the upper $T_J$ limit for JFCs exist. We adopt the upper limit for JFCs of $T_J<3.05$ from \citet{Gladman2008}.  By contrast, \citet{Jewitt2015-Asteroids4} set the dividing line between Jupiter-family comets and active asteroids, some of which are main-belt comets, at $T_J = 3.08$.
}

\textcolor{black}{\citet{Levison1996} defined ``Encke-type comets'' (ETCs) as comets with $a < a_J$ and $T_J > 3$ to distinguish ``decoupled'' comets like 2P/Encke, whose aphelion distance is $\approx 1$~au interior to Jupiter's orbit, from typical JFCs that undergo close encounters with Jupiter. We do not find this category useful, as 
objects defined this way are a mix of asteroids and comets, as we discuss below.
We consider Encke, which has $T_J = 3.025$, to be a JFC on an unusual orbit.}

\textcolor{black}{
Encke's perihelion distance $q = 0.34$~au (smaller than Mercury's semimajor axis) is unique among comets with $T_J > 3$. Encke is generally assumed to originate in the Kuiper belt, but numerical simulations that reproduce the comet's decoupled orbit require it to remain active for an implausibly long time  \citep{Valsecchi1995, Fernandez2002, Levison2006ETC}.
}

\textcolor{black}{
About half of the other ``Encke-type comets''  listed in the JPL Small-Body Database are active asteroids (see chapter by Jewitt \& Hsieh). The other half are JFCs that do not cross Jupiter's orbit, but still come within 2 Hill radii of the planet and interact strongly with it. 
Many of these JFCs are quasi-Hilda comets \citep{Tancredi1990, DiSisto2005, Toth2006, Ohtsuka2008, Gil-Hutton2016}, which undergo slow, and thus strong encounters with Jupiter, sometimes leading to temporary captures by the planet. 
Indeed, D/1993 F2 (Shoemaker–Levy 9) may have been a quasi-Hilda before its capture by Jupiter and impactful demise \citep{Chodas1996}. 
}

\textcolor{black}{Halley-type comets (HTCs) were once defined as periodic comets with orbital periods between 20 and 200 years \citep{Carusi1987b}, but now are typically classified as periodic comets with $T_J<2$ \citep[e.g.][]{Levison1994, Levison1996}. 
HTCs generally have larger semimajor axes (and higher inclinations) than the JFCs, but the new classification includes ``sun-skirting'' comets like 96P/Machholz with $a \approx 3$~au and
perihelion distances well within the orbit of Mercury \citep{Jones2018}. 
}

\textcolor{black}{
Halley-type comets have often been taken to be the shortest-period tail of the distribution of returning Oort Cloud comets. However, the population of known HTCs is small, about 10\% that of JFCs, in large part because HTCs' (generally) infrequent appearances provide few opportunities for discovery. The orbital distribution of Halley-type comets is therefore not very well constrained \citep[e.g.][]{Wiegert1999}. 
HTCs can be produced both from the scattering TNO population \citep[e.g.][]{Brasser2012, Levison2006} and from the Oort Cloud \citep[e.g.][]{Wang2014,Nesvorny2017}. More discussion of Encke- and Halley-type comets can be found in the Kaib \& Volk chapter.
}

The population of Centaurs, small bodies with orbits in the giant planet region, provides the link between the outer solar system TNO populations and the inner solar system short-period comets.
Most Centaurs are former TNOs that evolved onto Neptune-encountering orbits and were scattered inward to the giant planet region.
There they have relatively short dynamical lifetimes of $\sim$1--10~Myr \citep{Tiscareno2003, Horner2004, DiSisto2007, Bailey2009, Fernandez2018} and either evolve inward onto short-period comet orbits or outward onto highly eccentric orbits in the outer solar system; the details of this evolution are discussed in Section~\ref{sec:cent-to-JFCs}.
Cometary activity has been detected on some Centaurs with perihelion distances as large as $\approx 12$~au (e.g. \citealt{Jewitt2009, Lilly2021}; see discussion in Section~\ref{sec:centaur-activity}), further demonstrating that they are the transition population between the TNOs and JFCs. However, the presence of activity does not by itself show that the populations are related. For instance, although some active asteroids have sublimation-driven activity and are thus {\em bona fide} comets, most are thought to have resided in the asteroid belt for billions of years. 

\section{Origins of Trans-Neptunian Objects and Ecliptic Comets}
\label{sec:origins}

\subsection{Dynamical Origins of Trans-Neptunian Objects}
\label{sec:origins_tnos}

How was the Kuiper Belt populated? The belt is thought to be a
relic of an ancient massive planetesimal disk outside the orbits of
the giant planets \citep{Nesvorny2018}. \citet{Fernandez1984} showed that gravitational interactions of the giant planets with small bodies cause Jupiter to lose angular momentum (because it ultimately ejects most such bodies from the solar system) and Saturn, Uranus, and Neptune to gain angular momentum. If the planetesimal disk is massive enough (comparable to Uranus or Neptune), these interactions can change the planets' orbits substantially. \citet{Malhotra1993, Malhotra1995} showed that the outward planetesimal-driven migration of Neptune  would result in capture of Pluto and ``Plutinos'' into the 3:2 MMR. 
\textcolor{black}{
This model predicted that TNOs would also be captured into other MMRs, notably the 2:1 resonance near 48~au. TNOs in the 2:1 and many other MMRs were soon found by surveys such as DES \citep{Elliot2005}, CFEPS \citep{Petit2011}, and OSSOS \citep{Bannister2018} [see Section~\ref{sec:kuiperpops}].
}

In these migration scenarios, the original orbits of the giant planets would have been more closely spaced than they are now. 
\textcolor{black}{In the Nice model \citep{Tsiganis2005}, Jupiter and Saturn are initially closer together then cross a mutual 2:1 resonance as a result of interactions with the planetesimal disk, exciting their eccentricities and triggering an instability in the orbits of all the giant planets \citep{Tsiganis2005, Levison2008, Brasser2009, Morbidelli2010, Levison2011, Nesvorny2012, Izidoro2016, Gomes2018, Quarles2019, Nesvorny2021-tp}. }
More recent models of planetary instabilities have addressed problems with the Nice model, such as the excitation of excessively large eccentricities and inclinations in the terrestrial planets \citep{Brasser2009, Agnor2012}. Efforts to match the intricate orbital distributions of the trans-Neptunian populations, asteroid belt, and other small body reservoirs have also led to substantial revision of the Nice model (see reviews by \citealt{Dones2015,Nesvorny2018,Malhotra2019,Morbidelli2020,Raymond2022}). For instance, most current models invoke a fifth giant planet that was later ejected from the solar system by Jupiter \citep{Nesvorny2011, Nesvorny2012, Batygin2012, Cloutier2015} and find that the instability probably took place early, in the first 10--100~Myr of the solar system \citep{Deienno2017, Mojzsis2019, DeSousa2020}, rather than after hundreds of Myr \citep{Gomes2005}. 
\textcolor{black}{Our evolving understanding of the timing of the instability has implications for the pre-migration evolution of the disk, including the likely collisional histories of the planetesimals that eventually become the TNOs that feed the short-period comet populations \citep[e.g.][]{Morbidelli2015}.}

During the planetary instability, most bodies in the proto-trans-Neptunian disk are ejected from the solar system, but some reach quasi-stable niches, where they survive to the present. 
Models predict that the Oort Cloud and TNO populations contain the overwhelming majority of the survivors. 
Based on the known populations of small bodies, particularly Jupiter Trojans, and the capture efficiency found by modeling the instability, the disk is estimated to have originally contained about $6 \times 10^9$ bodies with diameters $d \gtrsim 10$~km \citep{Morbidelli2009, Nesvorny2018}. 
The instability model of \citet{Nesvorny2019-ossos}, evolved to the present, predicts the number of Centaurs of this size that OSSOS discovered to within a factor of two (also see \citealt{Bottke2022}). 
However, we know little about how many such objects reside in the TNO populations and especially the Oort Cloud.  
The fraction of those objects that reach the Oort Cloud and TNO populations and survive today are estimated to be $\eta = 0.05 \pm 0.01$ \citep{Vokrouhlicky2019} and 0.004 \citep{Nesvorny2017}, respectively, corresponding to $6 \times 10^9 \times 0.05 = 3 \times 10^8$ $d > 10$~km bodies in the Oort Cloud and $2 \times 10^7$ in the TNOs. 
Bodies from the disk are also implanted into the asteroid main belt ($\eta \approx 5-8 \times 10^{-6}$, \citealt{Levison2009, Vokrouhlicky2016}), as Hildas and Trojans of Jupiter (each with $\eta \approx (6 \pm 1) \times 10^{-7}$, \citealt{Nesvorny2013, Vokrouhlicky2016}), and as irregular satellites of Jupiter, Uranus, and Neptune (each with $\eta \approx 2--3~\times 10^{-8}$) and Saturn ($\eta \approx 5 \times 10^{-8}$, \citealt{Nesvorny2014}), yielding some 40,000 main-belt asteroids, 4,000 Hildas, 4,000 Trojans, and 100--300 irregular satellites of each giant planet.

\textcolor{black}{
Finding scenarios for the early evolution of the outer planets that are consistent with the complex orbital distribution of TNOs (Section~\ref{sec:kuiperpops}) is difficult and a topic of active research.
In general, some amount of planetesimal-driven migration \citep[e.g.][]{Fernandez1984,Malhotra1993,Malhotra1995} is invoked to explain the large number of resonant TNOs, while some sort of dynamical instability amongst the giant planets \citep[e.g.][]{Tsiganis2005} is invoked to help explain both the planets' orbits and help dynamically excite the TNOs.}
Three main migration models have been proposed, which depend on the value of
Neptune's eccentricity $e_{\rm N}$ at that time: 
(A) very low-eccentricity ($e_{\rm N} \lesssim 0.01$) migration of Neptune from $<25$~au to 30~au \citep[e.g.][]{Malhotra1993, Malhotra1995, Hahn2005}; 
(B) instability-driven scattering of Neptune from $<20$ au to an eccentric orbit ($e_{\rm N} \sim 0.3$) at $\sim 30$ au, and subsequent circularization of Neptune's orbit by dynamical friction from the planetesimal disk \citep[e.g.][]{Tsiganis2005, Levison2008}; 
and (C) an intermediate case in which Neptune's migration is interrupted by the instability, with its eccentricity reaching a peak of $e_{\rm
N}=0.03$-0.1, and then dropping to $\simeq 0.01$ as Neptune slowly
migrates toward 30 au \citep[e.g.][]{Nesvorny2012, Nesvorny2016}.
\textcolor{black}{As discussed below, these different kinds of migration models result in different distributions of the TNO populations that feed into the current comet populations.}

The \textit{original} mass and radial span of the Kuiper belt are unknown, but models that initially put several Earth masses beyond 30~au run into problems with Neptune's migration, because Neptune continues to migrate past that distance \citep{Gomes2004}. 
Migration beyond 30~au can also result in excess mass in the residual belt; today's TNOs with $r<50$ au represent $<0.1$ $M_{\oplus}$ \citep{DiRuscio2020}. 
A sharp truncation of the original massive disk, an exponential cutoff of the disk surface density near 30 au, or some combination of the two is required to match Neptune's orbit \citep{Nesvorny2020-colors}. 
Specifically, the disk's mass density at 30 au must have been $\lesssim 1 M_{\oplus} {\rm au}^{-1}$ for Neptune to stop at 30 au \citep{Nesvorny2018}. 
The original disk most likely continued, with a low surface density, to 45 au, where the cold classical TNOs formed and survived \citep{Batygin2011}. 
\textcolor{black}{
The current mass of cold TNOs is estimated to be only $\sim 3\times10^{-4}$ $M_{\oplus}$ \citep{Fraser2014} to $(3 \pm 2) \times 10^{-3}$ $M_{\oplus}$ \citep{Nesvorny2020-colors}. 
These values assume that TNOs have densities of 1 g cm$^{-3}$.}
\textcolor{black}{The vast majority of the dynamically hot TNO populations that feed the Centaur and short period comet populations originate in the more massive, closer-in portion of this original planetesimal disk.}

\begin{figure}[!htb]
\begin{center}
\includegraphics[width=0.49\textwidth]{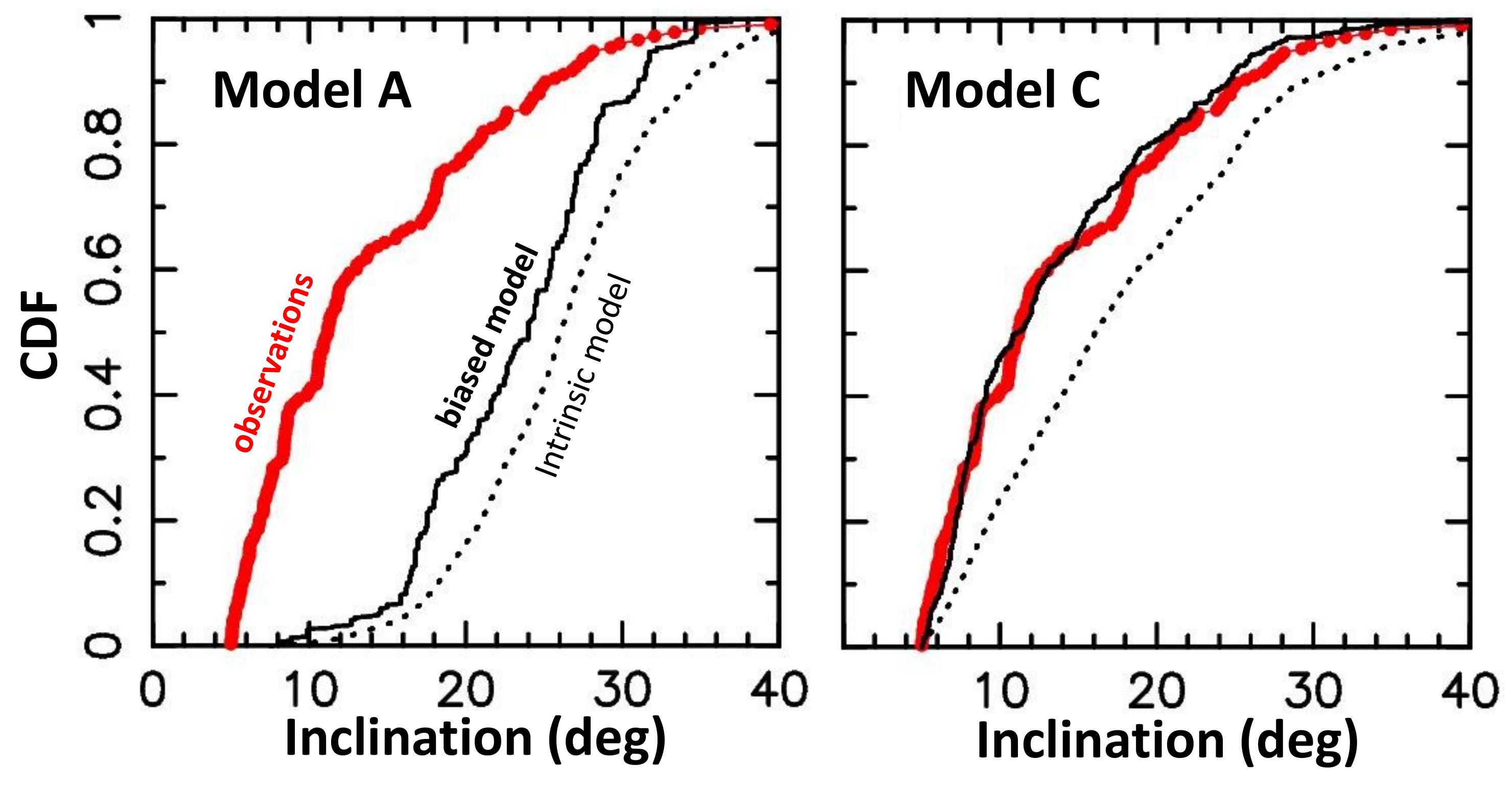}
\caption{The cumulative distribution function (CDF) of TNO orbital inclinations obtained in dynamical models A (left panel) and C (right panel). The intrinsic-model, biased-model and observed distributions are shown by dotted, solid and red lines, respectively. The observed distribution
shows all detections from the Outer Solar System Origins Survey (OSSOS) with $40<a<47$ au, $q>36$ au and
$i>5^\circ$ \citep{Bannister2018}. The biased model distribution was obtained by applying the OSSOS survey simulator \citep{Lawler2018-survey} to the intrinsic model. In model A, Neptune migrated from 24 au to 30 au with $e_{\rm N} \simeq 0.01$ on a 10-Myr timescale. In model C, Neptune's eccentricity was excited to $e_{\rm N} \simeq 0.1$ when Neptune reached 28 au and slowly damped after that such that $e_{\rm N} \simeq 0.01$ in the end. All planetesimals shown here started within 30 au of the Sun.}
\label{fig:nesvorny-models}
\end{center}
\end{figure}

\textcolor{black}{
The three models outlined above have different implications for the hot TNO populations. 
A pure instability model (model B) does not explain the wide orbital inclination distribution of hot TNOs \citep{Petit2011, Nesvorny2015} because the instability happens so fast that there is not enough time to sufficiently excite the orbital inclinations. 
The existence of cold classical TNOs also limits how large Neptune's eccentricity can be when it reaches its current semimajor axis ($e_{\rm N}<0.15$; \citealt{Dawson2012}).
In all three models, planetesimals originating from the $<30$~au portion of the disk are scattered by Neptune to higher-$a$ and high-$e$ orbits; to end up in today's metastable hot TNO populations, these orbits must subsequently be stabilized by some dynamical mechanism that causes their orbital eccentricities to drop. }
The Kozai resonance \citep{Kozai1962} near and inside mean-motion resonances with Neptune is presumably the dominant implantation mechanism during periods of slow migration (\citealt{Nesvorny2020}). 
The Kozai resonance produces anti-correlated oscillations of $e$ and $i$. As the TNO's orbital eccentricity decreases, its orbit can decouple from Neptune scattering (i.e., the TNO's perihelion distance can evolve to beyond Neptune's aphelion distance) or drop out of resonance while the TNO's orbital inclination increases. 
The $\nu_{18}$ secular resonance \textcolor{black}{(which occurs when a TNO's orbital nodal precession rate matches one of the dominant nodal precession rates of the giant planets, and is the mode most associated with Neptune's nodal rate)} can also influence the inclination distribution \citep{Volk2019}, but it does not wipe out the Kozai signature.
\textcolor{black}{
When the Kozai implantation dominates for smooth migration (model A), it is difficult to obtain implanted hot TNO orbits with $i<10^\circ$ 
(Figure~\ref{fig:nesvorny-models}).}
\textcolor{black}{
The finer details of these stabilization mechanisms and how they affect the efficiency of implantation from the original planetesimal disk into the hot TNO populations is still not fully explored, even in the simplest planetesimal driven migration scenario (model A; see discussion in \citealt{Volk2019}).
}

\textcolor{black}{
In general, the results of migration models better match the observed hot TNO populations if Neptune's orbit is modestly excited by an instability at some point during migration (model C). 
This is particularly true for matching the inclination distribution of the hot TNOs (Figure~\ref{fig:nesvorny-models}).
}
In this case, bodies are implanted in the hot TNO populations as the eccentricities of high-$a$ and high-$e$ orbits drop due to the $\nu_8$ secular resonance \textcolor{black}{(which occurs when a TNO's perihelion precession rate matches one of the dominant perihelion precession rates of the giant planets, the mode most associated with Neptune's perihelion precession rate)}, which is stronger in this case, because $e_{\rm N} \neq 0$ \citep{Nesvorny2021}; 
here the contribution of Kozai cycles is relatively minor. 
As the $\nu_8$ resonance does not affect inclinations, the inclination distribution is roughly preserved during implantation.
The inclination distribution is then primarily controlled by Neptune's
\textcolor{black}{
scattering and the $\nu_{18}$ secular resonance; in some simulations, the slower Neptune's migration is, the
broader the inclination distribution becomes \citep{Nesvorny2015}, although this is not always a generic outcome of slow migration simulations \citep{Volk2019}. }
This model, assuming that Neptune's migration was long-range and slow 
($e$-folding time $\gtrsim 10$ Myr) better matches the Kuiper Belt's orbital structure, such as its inclination distribution
(Figure~\ref{fig:nesvorny-models}) and the fraction of TNOs in resonant orbits \citep{Nesvorny2018}.

The scattered disk is thought to be the main source of ecliptic (or
Jupiter-family) comets \citep{Duncan1997, Levison1997,
Volk2008, Brasser2013, Nesvorny2017} (see Section~\ref{sec:cent-to-JFCs} and \citealt{Dones2015} for a review of this topic). 
The dynamical structure of the inner detached disk (50--100 au), with
dropout bodies -- objects that fell out of resonance during migration on the sunward side of mean motion resonances
\citep{Bernardinelli2022} -- is an important constraint on Neptune's
migration \citep{Kaib2016, Nesvorny2016-beyond-50, Lawler2019}. 
\textcolor{black}{The efficiency of implantation into these populations and how the inner detached disk may feed the actively scattering population on long timescales depends on the details of migration.}
The scattered disk decayed by a factor of $\sim100$--300 since its
formation 4.5 Gyr ago (see, e.g., the scattered disk/Oort Cloud formation and evolution models of \citealt{Brasser2013} and \citet{Nesvorny2017}). 
This is reflected by a rapidly decreasing number of ecliptic comets and planetary impactors soon after the instability, and a gradual decrease during the past $\approx 4$~Gyr.

\subsection{Dynamical Routes from the Centaurs to the JFCs} 
\label{sec:cent-to-JFCs}

\begin{figure}[htp]
    \centering
    \includegraphics[width=0.49\textwidth]{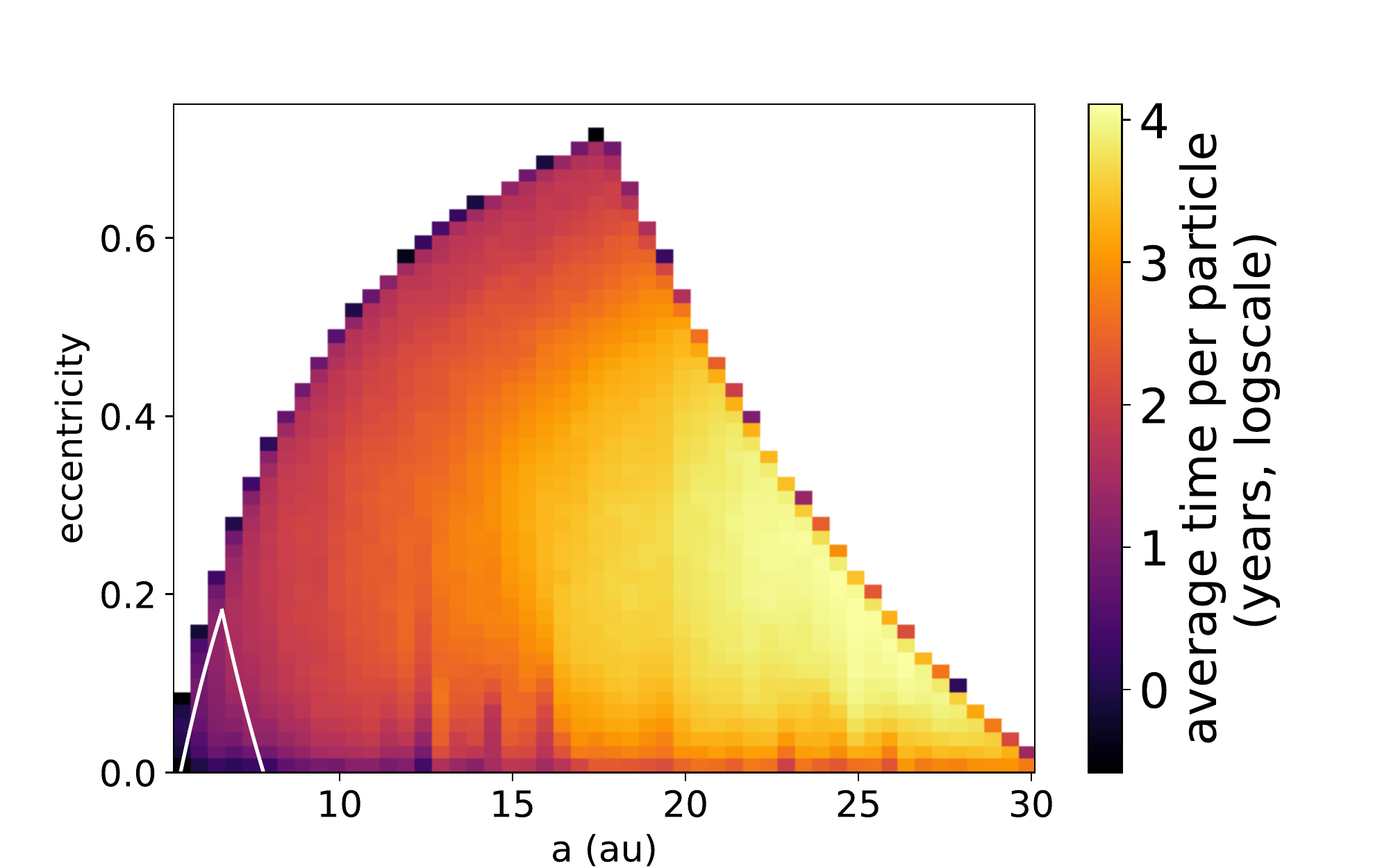}
    \caption{The colormap shows the time-weighted orbital distribution of Centaurs (using a definition whereby Centaurs' orbits are entirely enclosed in the Jupiter-Neptune region) in a simulation of their evolution from the outer solar system into the JFC population (figure from \citealt{Sarid2019}).}
    \label{fig:Centaur}
\end{figure}

The short-period comets are fed into the inner solar system from the Neptune-crossing population of TNOs. 
This is predominantly the population of scattering objects (see discussion in the chapter by Kaib \& Volk), though other sub-populations of TNOs contribute as well (see Section~\ref{sec:tno_sources}). 
Some objects on Neptune-crossing orbits will be scattered inwards into the Centaur population, from which point their evolution is dominated by gravitational scattering by the giant planets.
\textcolor{black}{Objects can spend upwards of $\sim100$~Myr in the scattering TNO population with perihelion very near or interior to Neptune.}
Once an object has been transferred onto an orbit in the giant planet region, it takes $\sim$~1- 10~Myr for that object to either be ejected back into the scattering population, or transferred into the JFC population \citep[e.g.][]{Tiscareno2003,DiSisto2007,Bailey2009,Sarid2019,DiSisto2020}.

Figure~\ref{fig:Centaur} illustrates the typical timespan objects spend traversing each part of the Centaur region \textcolor{black}{(which in this figure from \citealt{Sarid2019} is defined as an orbit entirely enclosed between Jupiter and Neptune, $q>5.2$~au and $Q<30.1$~au, which is slightly more restrictive than the \citealt{Gladman2008} definition we typically use).} 
Most of the time an object spends in the Centaur population is spent in the Uranus-Neptune region because encounters with these planets are gentler than those with Jupiter and Saturn. 
Even so, 20\% of Centaurs will be ejected back into the outer solar system before making it inside Uranus's orbit; only roughly half the Centaurs will be transferred onto orbits inside Saturn's, with roughly one third making it past Jupiter to become JFCs (\citealt{Sarid2019}; see also, e.g., \citealt{Tiscareno2003,DiSisto2020}).
Of order 1\% of Centaurs will end their journeys via impact with one of the giant planets \citep[e.g.,][]{Levison2000, Tiscareno2003, Raymond2018, Wong2021}.
Dynamical evolution in the Centaur region is relatively insensitive to the exact source population in the outer solar system, though Centaurs with high orbital inclinations do tend to have longer lifetimes in the giant planet region due to the decreased probability of close encounters with the planets \citep[e.g.,][]{DiSisto2020}.
Very high inclination and retrograde Centaurs are not well-explained by the observed populations of TNOs described in Section~\ref{sec:kuiperpops} because the journey through the giant planet region does not radically alter the inclination distribution of Centaurs relative to their source population \citep[e.g.,][]{Brasser2012,Volk2013}.
The observed high-inclination Centaurs and scattering objects \citep[e.g.,][]{Gladman2009,Chen2016} likely originate from non-TNO source regions (see discussion in the chapter by Kaib \& Volk).

The approximately one-third of Centaurs that become JFCs \citep{Levison1997, Tiscareno2003, DiSisto2007, Bailey2009, Fernandez2018} typically transfer into the inner solar system via a low-eccentricity orbit just exterior to Jupiter like that of comet 29P/Schwassmann-Wachmann 1; this orbital region, dubbed the JFC Gateway \citep{Sarid2019}, represents an important phase in the transition from Centaur to JFC because it coincides with thermal conditions that allow for significant cometary activity from sublimation of water ice (see, e.g., \citealt{Steckloff2020}). 
The Centaur to JFC transition is, of course, reversible, with objects passing back into the Centaur population (also typically through a Gateway orbit). 
How many JFCs survive, and in what state, to re-enter the Centaur population remains an open question. 

Dynamical models of JFC orbital evolution indicate that, after becoming JFCs, it can take up to $\sim0.5$~Myr for a comet to be dynamically removed \citep[e.g.,][]{Levison1994, DiSisto2009}.  
However, during this extended orbital evolution, the JFCs' orbital inclinations are pumped up to larger and larger values, driving the overall simulated JFC inclination distribution to disagree with the observed one \citep[e.g.,][]{Levison1994,DiSisto2009, Nesvorny2017}.
\textcolor{black}{
To solve this mismatch between simulated and observed JFC populations, it is usually assumed that a comet becomes inactive after some number of orbits with a perihelion distance below some threshold (often $q\lesssim 2.5$~au). 
Fits to the inclination distribution of JFCs with diameters between 1 and 10~km suggest physical lifetimes of $\approx$~4000--40,000 years ($\approx$~500-5000 orbits), which is a small fraction of the dynamical lifetimes \citep[e.g.][]{Levison1997,DiSisto2009, Brasser2015,Nesvorny2017}. Sub-km JFC nuclei must have much shorter physical lifetimes.
The exact nature of this fading (i.e., whether it is due to buildup of a surface lag layer (``mantle'', e.g. \citet{Rickman1990}) or physical destruction of small comet nuclei through spin-up or other mechanisms) is not well-understood; observations of dormant comet nuclei in various solar system regions (including the Centaur population) could help constrain this problem. 
An additional complication is that the activity of JFCs can turn on and off repeatedly. 
For instance, comet Blanpain (now known as 289P) was discovered in 1819, observed for two months, and not seen again for almost two centuries until it was identified with the asteroid 2003 WY25 and found to have a weak coma 
\citep{Jewitt2006}. 
}

\subsection{Kuiper Belt Sources of Centaurs and Comets} 
\label{sec:tno_sources}

\begin{figure}[!htb]
    \centering
    \includegraphics[width=0.49\textwidth]{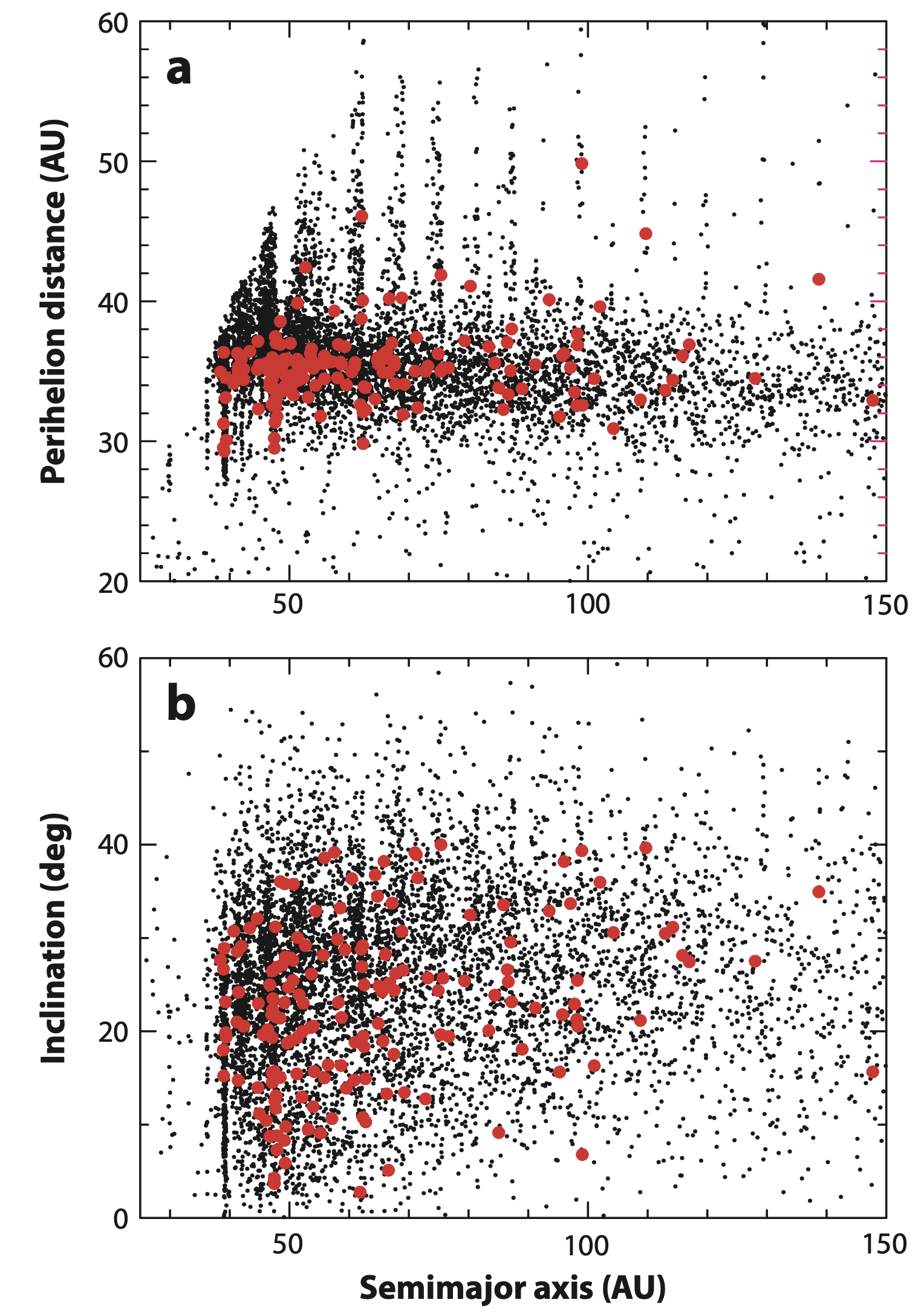}
    \caption{A model of the TNO population (modeled from formation to the present day) with the source-region orbits of eventual ecliptic comets highlighted.
    The black dots show a snapshot of the TNO population from $\sim3$ Gyr ago; objects from that snapshot that later become ecliptic comets (defined as $2 < T_J < 3$, $P <$~20 years, $q < 2.5$~au) are shown as larger red dots.
    About 75\% of TNOs that become short-period comets in this simulation originate in the scattered disk, with semimajor axes between 50 and 200~au; $\sim20$\% originate from TNOs with $a < 50$~au (some of which enter the 50-200~au scattering population en route); about 3\% come from the Oort Cloud.  
    From \citet{Nesvorny2018}, which was adapted from \citet{Nesvorny2017}.}
    \label{fig:TNO-JFC-source}
\end{figure}

\begin{figure}[!ht]
    \centering
    \includegraphics[width=0.49\textwidth]{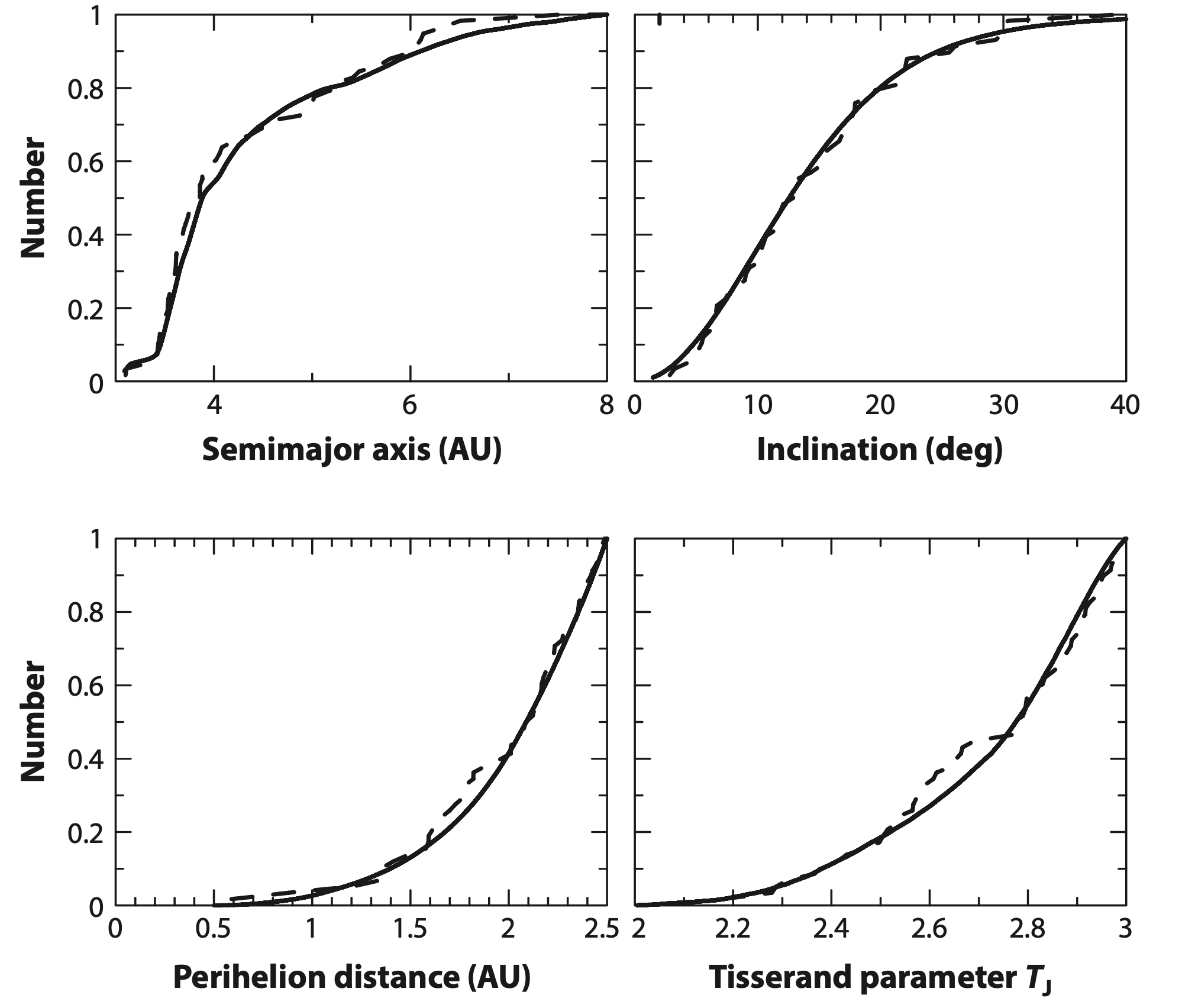}
    \caption{The cumulative orbital distributions of ecliptic comets with orbital periods $P < 20$~ years, Tisserand parameter with respect to Jupiter $2 < T_J < 3$, and perihelion distance $q < 2.5$~au. The model results (solid lines; \citealt{Nesvorny2017}) are compared with the distribution of known JFCs (dashed lines). The model assumes that ecliptic comets remain active and visible for $N_p(2.5) = 500$ perihelion passages with $q < 2.5$~ au. From \citet{Nesvorny2018}, which was adapted from \citet{Nesvorny2017}. The panels are normalized so that the numbers of ecliptic comets with $q < 2.5$ and with $2 < T_J < 3$ equal one.}
    \label{fig:ecliptic-comets}
\end{figure}

Dynamical models of the evolution of trans-Neptunian populations confirm that the dynamically hot populations, particularly the scattering population, dominate the influx of new Centaurs and then JFCs in the inner solar system.
Figure~\ref{fig:TNO-JFC-source}, for example, highlights the original orbits of particles that evolve onto cometary orbits from a model of the TNOs \citep{Nesvorny2017}. Most of these comet-supplying TNO orbits have perihelion distances below $\sim37-38$~au, placing them in the scattering population, the one originally argued to be the dominant supplier of JFCs \citep[e.g.][]{Duncan1997}. 
However, it is clear that a few other TNO populations also contribute to the influx.
Figure~\ref{fig:TNO-JFC-source} shows a number of JFCs sourced from Neptune's resonant populations (vertical features most evident in the top panel). 
Most notable are the close-in 3:2 and 2:1 resonances, which are known to have large populations \citep[e.g.][]{Volk2016}; chaotic diffusion within these resonances (and others) can slowly feed objects onto Neptune-crossing orbits that supply the JFCs \citep[e.g.][]{Morbidelli1997,Tiscareno2009}. 
Figure~\ref{fig:TNO-JFC-source} also shows a handful of high-perihelion TNO orbits that nonetheless evolve into JFCs.
Most of these simulated objects appear to be in Neptune's more distant mean motion resonances, where secular effects cause their perihelia to cycle in and out of Neptune's reach \citep[e.g.][]{Gomes2005kz}.
In the real TNO population, it is still not entirely clear how many of the so-called `detached' TNOs on high-perihelion orbits are still in Neptune's resonances (and thus could cycle back into the scattering population) and how many are truly stranded at high perihelia (for example, by being dropped out of resonance during the late stages of Neptune's migration; e.g., \citealt{Kaib2016,Lawler2019}); 
this is because the orbits of very large semimajor axis TNOs are difficult to measure to high-enough precision for this determination (discussed in, e.g., \citealt{Gladman2021}).
The orbital distribution of the detached TNOs and how much that population can cycle in and out of the actively scattering population is one of the larger uncertainties in determining the exact delivery rate of JFCs from the trans-Neptunian region (also relevant to this are the effects of any unknown perturbers in the distant solar system; see discussion in \citealt{Nesvorny2017}).
However, following the evolution of our best models of the TNO populations into the inner solar system \citep{Levison1997,DiSisto2009, Brasser2013, Nesvorny2017} results in an excellent match to the observed orbital distribution of comets (Figure~\ref{fig:ecliptic-comets}), confirming that the majority of JFCs likely originate as TNOs.

For completeness, we note that a few other solar system populations may contribute to the observed JFCs. 
These include the low-inclination Themis family in the outer main asteroid belt 
\citep{Hsieh2020}  (also see the chapter by Jewitt \& Hsieh), the Hildas \citep{DiSisto2005} in the 3:2 resonance with Jupiter near 4~au, and the Jupiter and Neptune Trojans \citep{horner_lykawka_2010,DiSisto2019}.
These sources are likely minor suppliers of the JFC population compared to the TNOs. Bodies from the primordial Kuiper Belt were implanted into all these populations (see Section~\ref{sec:origins_tnos}), so JFCs from different present-day sources may not show clear signs of their birthplaces. \textcolor{black}{Indeed, two large asteroids near the middle of the asteroid belt, 203 Pompeja and 269 Justitia, have recently been proposed to have originated as TNOs, based on their colors, which are even redder than D-type Trojans and Hildas \citep{Hasegawa2021}.}

\section{Compositions and Physical Properties}
In this section, we discuss the compositions of TNOs, Centaurs, and JFCs as observed from Earth- and spacecraft-based telescopic observations.  Measured properties of these objects are valuable for constraining solar system formation models \citep{Barucci2011,vanderWiel2014}. 
As Centaurs are thought to be an intermediate stage in the orbital evolution from TNOs to JFCs, intercomparison of their compositions can provide insights into how the journey from the scattering TNO population inward influences outgassing behavior and observed coma composition. Where available, we discuss the few compositional links between these dynamically connected populations.

\subsection{Surface Compositions of TNOs and Centaurs}
\label{sec:kbo_ices}

\begin{figure}[!ht]
\begin{center}
\includegraphics[width=0.49\textwidth]{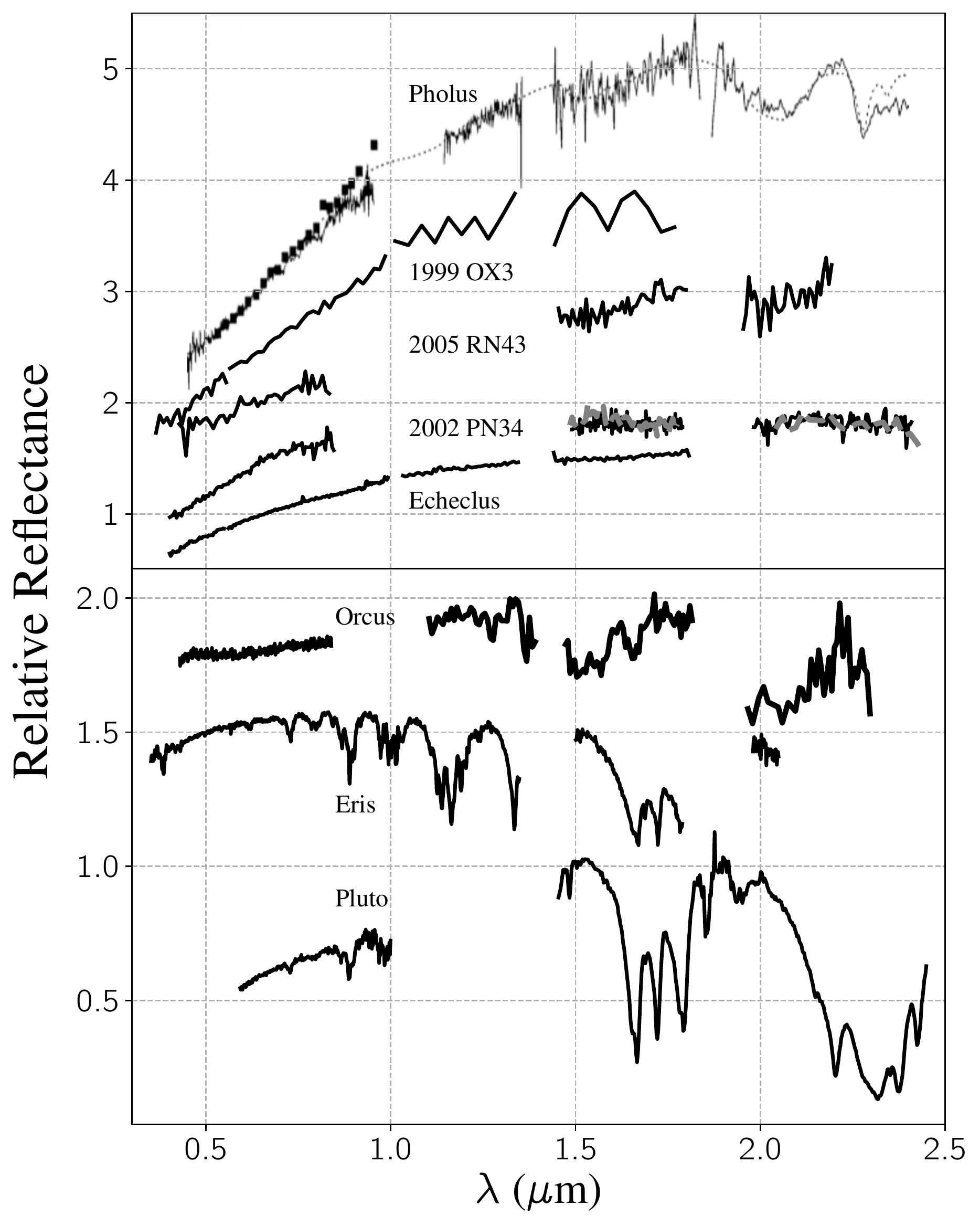}
\caption{Select TNO and Centaur reflectance spectra. 
\textbf{Top:} spectra of (from top to bottom) the inactive Centaur 5145 Pholus \citep{Cruikshank1998}, TNOs (44594) 1999 OX3 \citep{Seccull2018}, (145452) 2005 RN43 \citep{Alvarez-Candal2008,Guilbert2009}, and (73480) 2002 PN34 (black \citealt{DeMeo2010}; gray \citealt{Brown2012}), and the active Centaur 60558 Echeclus, also known as comet 174P \citep{Seccull2018}. 
Spectra were normalized to unity at $0.65 \mbox{ $\mu$m}$ and offset vertically in steps of 0.45. 
The NIR spectrum of 2005~RN43 has been renormalized to the optical spectrum using its observed (V-R) color. 
The dashed curve is a spectral model of Pholus, which includes contributions from olivine, tholin, water ice, and methanol. 
\textbf{Bottom:} Spectra of the dwarf planets Orcus \citep[top;][]{DeMeo2010}, Eris \citep[middle;][]{Alvarez-Candal2011} and Pluto \citep[bottom;][]{Merlin2010}. The spectra were normalized to unity at $1.5 \mbox{ $\mu$m}$, and those of Orcus and Eris have been offset by 0.8 and 1.0 vertically. In both panels, regions affected by telluric lines or instrumental artifacts have been removed, and data have been rebinned for clarity. \textcolor{black}{ This figure omits notable objects for which the published spectra were not available to the authors, including that of the methane rich Quaoar \citet{Jewitt2004}, and the water-ice rich Haumea \citet{Trujillo2007}.}} \label{fig:TNOSpectra}
\end{center}
\end{figure}

Beyond atmospheric escape through processes such as Jeans loss \citep{Lisse2021}, TNOs are generally inactive bodies. As such, specific material identification on surfaces is made entirely through reflectance spectroscopic remote sensing. TNOs tend to have spectra that are largely devoid of identifying features (Figure~\ref{fig:TNOSpectra}). The detection of specific materials on the surfaces of TNOs and Centaurs is so scarce that we can \textcolor{black}{easily tabulate them in Table~\ref{tab:moleculesseen}, which shows the limited list of surface ices detected on these objects.}

\begin{table*}
\centering
\caption{Representative listing of molecules detected/inferred on surface ices and in comae or atmospheres (volatiles) of TNOs, Centaurs, and comets}  \label{tab:moleculesseen}
\begin{tabular}{|l l l l l l l l l |}
 \hline
TNO surface ices & H$_2$O & CH$_3$OH & CH$_4$ & N$_2$ & CO & CO$_2$ & NH$_3$ & C$_2$H$_6$ \\ 
TNO volatiles & N$_2$ & CO & CH$_4$ & & & & & \\
\hline
Centaur surface ices & H$_2$O & CH$_3$OH & & & & & & \\
Centaur volatiles & CO & H$_2$O & CO$_2$ & HCN & CN & CO$^+$ & N$_2^+$ & \\
\hline
Comet ices & H$_2$O & CO$_2$ & COOH-group & & & & & \\
Comet volatiles  & H$_2$O & CO$_2$ & CO & CH$_4$ & HCN & CH$_3$OH & H$_2$CO & NH$_3$ \\
& HNC & C$_2$H$_2$ & C$_2$H$_6$ & HCOOH & HCOOCH$_3$ & HNCO & H$_2$S & OCS  \\
& HC$_3$N & SO & SO$_2$ & CS & CH$_3$CO & S$_2$ & CN & C$_2$ \\
& NH$_2$ & C$_3$ & CO$^+$ & N$_2^+$ & H$_2$O$^+$ & CO$_2^+$ & N$_2$ & \\

 \hline
\hline

\end{tabular}
\flushleft

\endflushleft

\end{table*}

The most frequently detected material on TNOs is water ice, in both amorphous and crystalline forms \citep{Zheng2009,Barucci2011}. Water ice is thought to be abundant in all TNOs and Centaurs \citep{Brown2012}, and it is clearly one of the major building blocks of \textcolor{black}{those bodies. Water ice absorption is most clearly seen in Haumea and its collisional family \citep{Barkume2006, Trujillo2011}.} Methanol was first detected on the surface of the Centaur Pholus \citep{Cruikshank1998}, and has been detected with varying levels of confidence on only a handful of TNOs through ground-based spectroscopy \citep{Barucci2011}, and in the spectrum of Arrokoth by the New Horizon spacecraft \citep{Grundy2020}. 
Its rarity of detection implies either that its spectral signatures are usually too weak to detect, either because it is not common on TNO surfaces, or because those signatures are masked by other materials. The former condition might imply that methanol is a by-product of some post-formation process, rather than a material that was abundant during the formation of TNOs. 

The remaining materials are only found on the largest TNOs, i.e., on ``dwarf planets''\footnote{We caution that dwarf planets should not be conflated with  volatile-rich TNOs. For an object to be considered a dwarf planet requires only that it be massive enough for self-gravity to govern its shape. It is a coincidence of their formation and current surface temperatures that the dwarf-planet TNOs are the only bodies able to retain detectable levels of the volatile ices, which is the condition we are interested in here.}, such as Eris, Pluto, Makemake, and Quaoar \citep[for a recent summary, see][]{Barucci2020}. 
These are the known volatile ices at TNO temperatures ($\approx$~30-50 K) that are unstable to sublimation on timescales shorter than the age of the solar system: N$_2$, CH$_4$, CO, and CO$_2$. It is only by virtue of the masses of the largest TNOs that atmospheric retention of these ices is sufficient to preserve a detectable abundance of these materials since formation \citep{Schaller2007,Brown2011,Lisse2021}. At the most basic level, the interplay between temperature and gravity governs the retention of each volatile, resulting in dramatic variations in relative abundances from object to object. This explains why some objects have spectra that are N$_2$ dominated (e.g. Eris), some are methane dominated (e.g. Quaoar), and some exhibit features of all the above volatile ices (e.g. Pluto). Figure~\ref{fig:TNOSpectra} presents a small sample of spectra of the dwarf TNOs, exhibiting the sheer diversity of their spectra.
Much like water on the Earth, the presence of these materials in a semi-stable state drives interesting atmospheric and surface re-circulation cycles that dominate the surface rheologies of these bodies \citep{Bertrand2018,Hofgartner2019}. 
These processes and the loss rates, which are highly sensitive to the surface and interior temperature histories of these bodies, greatly complicate estimates of the primordial ice abundances. This interesting topic is well beyond the scope of this chapter. For our purposes it is sufficient to emphasize that these volatile ices are primordial in nature, and are not fully depleted from the interiors of smaller TNOs, the interior presence of which is betrayed by the presence of these ices in the comae of JFCs (see Section~\ref{sec:Centaur_comps}).

Of course, the compositions of TNOs are not limited only to the few ices that have been spectrally identified. 
TNOs typically have red colors at optical wavelengths.
The most striking example of this is the New Horizons flyby imagery of the small (36 km long) TNO Arrokoth \citep{Grundy2020}. The dynamic range of TNO/Centaur optical colors is enormous, spanning from nearly neutral reflectors (e.g. the Haumea family members; \citealt{Brown2007}) to some of the reddest objects in the solar system (e.g., Gonggong - 2007 OR10; \citet{Fraser2012}). 
This red color is most commonly attributed to chemical processing of organic materials by cosmic rays \citep{Thompson1987,Cruikshank1998,Barucci2011,Fraser2012}.
Such materials exhibit the so-called optical gap absorption feature, driven by the C-H $\pi$-bond. Centered in the NIR, the shape and width of this feature depend on the amount of dehydrogenation of the organic, how disordered the molecular structure is, and the level of non-organic contaminant in the molecular chains, in particular \citep{Seccull2021b}. 
This feature is common to many organic materials that are considered suitable astrophysical analogs; compare the chemically simple polycyclic aromatic hydrocarbons \citep[e.g.,][]{Izawa2014} to the highly disordered laboratory materials called tholins \citep[e.g..][]{Roush2004}, both of which show this deep optical gap.
If attribution to organic materials is correct, the red colors of TNOs imply that simple organics such as methane were abundant for enough time in the early solar system that red dehydrogenated crusts could develop \citep{Brunetto2006} before those volatile ices were depleted.

The spectrum of Arrokoth reveals a mostly featureless spectrum typical of most TNOs, except for two weak features at 1.8 and $2.2\mbox{ $\mu$m}$. 
The absorption at $2.2\mbox{ $\mu$m}$ has been attributed to methanol ice \citep{Grundy2020}. 
It has been suggested that the $1.8\mbox{ $\mu$m}$ 
feature can be driven by the presence of sulfur in the organic residue \citep{Mahjoub2021}. 
Irradiation experiments on laboratory ice mixtures containing methane, ammonia, hydrogen sulfide, and water result in a red material bearing an absorption at $1.8 \mbox{ $\mu$m}$ that is absent in mixtures with no sulfur \citep{Mahjoub2021}.  
Such a feature has not been detected on any other TNO, possibly as a result of the low available signal-to-noise ratio of even the best spectra of small TNOs. It may be that the feature is unique to the so-called cold-classical TNOs (see Section~\ref{sec:origins_tnos}) which are the only TNO population presumed to have formed in-situ. The search for sulfur in the comae of JFCs has taken on new importance, as the presence or absence of sulfur-bearing materials in JFCs would provide significant insights regarding the compositional variations in the protoplanetesimal disk. The recent strong detection of atomic sulfur at 1425~{\AA} in the JFC 46P/Wirtanen rivals that of atomic hydrogen \citep{Noonan2021}, and is consistent with emission directly from the nucleus or from grains very near the surface, similar to atomic sulfur and other sulfur-bearing species in 67P/Churyumov–Gerasimenko \citep{Calmonte2016}. Additional searches are underway for other sulfur-bearing species in JFCs to test models of cometary ices and their subsequent processing history  \textcolor{black}{\citep{Presler-Marshall2020, Saki2020, Altwegg2022}}.

No discussion about TNO compositions would be complete without the mention of silicate materials. Two TNOs, the $\approx 700$ and 300-km diameter Plutinos (208996) 2003 AZ84 \citep{Fornasier2009} and (120216) 2004 EW95 \citep{Seccull2018}, exhibit spectra that appear similar to C-type asteroids, with absorption features consistent with hydrated silicates. 
These two bodies are not icy, lacking even the absorptions due to water ice at 1.5 and $2.0 \mbox{ $\mu$m}$ \textcolor{black}{seen in the Haumea family \citep{Brown1999}}. Frustratingly, silicate materials have avoided spectroscopic detection for icy TNOs. 
Their presence was first inferred by the need for silicate materials to account for the densities of the largest TNOs. Of the eleven TNOs with diameters $d > 600$~km and measured densities $\rho$, all but one, 55637 (2002 UX25), have $\rho >$ 1 g cm$^{-3}$ \citep{Brown2013, Brown2017} [see \citet{Bierson2019} and \citet{Grundy2019} for densities of other TNOs]. The largest density is $2.43 \pm 0.05$ g cm$^{-3}$ for Eris/Dysnomia, the most massive TNO system known \citep{Holler2021}.  The presence of silicates has also been inferred from spectroscopic modeling, either of individual spectra (see \citealt{Barucci2011}, for example), or of the continuum of TNO optical and NIR colors \citep[e.g.][]{Fraser2012}. Recent modeling of spectrophotometry spanning $\sim0.5-4.5 \mbox{ $\mu$m}$ implies the presence of silicate-rich surfaces amongst some of the brighter TNOs \citep{Fernandez-Valenzuela2021}. These results demonstrate that silicates \emph{do} exist in icy TNOs (of course they do!), but simply are masked by the presence of other materials. 
It is likely that characterization of silicates in TNOs will have to await observations at longer wavelengths ($\lambda\gtrsim3 \mbox{ $\mu$m}$; \citealt{Parker2016}). Spectral observations from JWST are likely to be quite important in this regard (see Section~5). 

Beyond spectral studies, many insights regarding the compositions of TNOs have come from broadband photometric techniques. We review those here. 

Significant effort has been devoted to developing a taxonomic system for TNOs, as even the simple act of determining the number of classes is likely to influence our interpretation of the early solar system. It is generally thought that the varied classes of TNOs reflect the compositional structure of the regions of the protoplanetary disk from which TNOs originated. Proposed mechanisms for compositional differences of planetesimals include ice lines \citep{DalleOre2015} and post-formation volatile loss \citep{Wong2016,Brown2011}, which would lead to variable composition with distance from the Sun, prior to disk dispersal by the migrating gas giants. It is important to highlight that to date, TNO compositional measurements are all measured from reflected light, and are typically assumed to be indicative of primitive nucleus values.

Efforts to determine the number of TNO classes have been frustrated by their nearly featureless spectra.  
TNO reflectance spectra \citep[e.g.][]{Fornasier2009,Guilbert2009,Barkume2008,Barucci2011}, like those of most icy bodies, including the Jupiter Trojans, Centaurs, and JFCs \citep{Emery2011, Cruikshank1998, Barucci2002} can be broadly described as linear in the optical and near-infrared (NIR), with a different spectral slope in each region, and a smooth roll-over or transition between the two slopes, very roughly centered at $\sim$0.9~$\mu m$, and sometimes an absorption band of water ice at 1.5~$\mu m$ and other ices in the infrared (see Sec. \ref{sec:kbo_ices}). A few examples are presented in Figure~\ref{fig:TNOSpectra}. 
By comparison, asteroids generally exhibit distinct features between classes that aid taxonomic interpretation \citep{DeMeo2009}; such features are generally unavailable for the classification of TNOs. Those few materials that have been confidently detected from the spectra of TNOs, such as water ice, 
do not seem to belong uniquely to certain TNO taxa. Rather, taxonomic systems have been largely generated from the variations in optical and NIR colors seen from object to object. 

Centaurs and most dynamically excited TNOs exhibit a bifurcated optical color distribution \citep{TeglerRomanishin1998, Peixinho2012,Tegler2016,Fraser2012,Peixinho2015,Lacerda2014,Marsset2019}. The bimodal color distribution has provided a functional taxonomy of  two populations for small TNOs, now colloquially referred to as Red and Very Red. In this basic system, Pluto is a Red member, and Arrokoth belongs to the Very Red population of TNOs \citep{Grundy2020}. Intrinsically, Red class TNOs outnumber the Very Red objects by at least 4-to-1 \citep{Wong2017,Schwamb2019}. This ratio has been used to infer that the purported compositional line that divides the proto-Red and proto-Very Red populations fell between 30 and 40 au \citep{Nesvorny2020-colors, Buchanan2022}.

\begin{figure}[ht!]
\begin{center}
\includegraphics[width=0.49\textwidth]{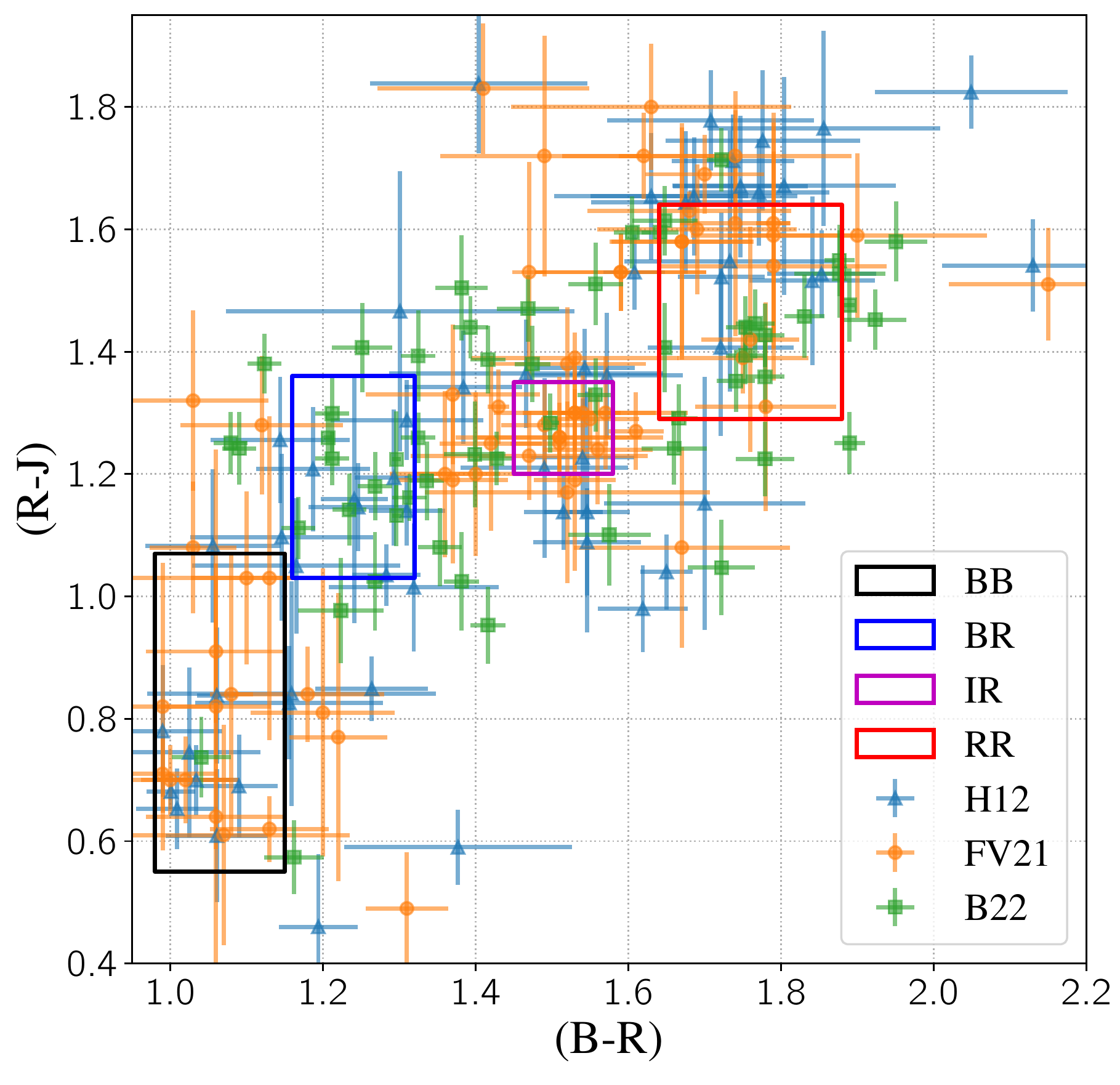}
\caption{
\textcolor{black}{Colors of TNOs and Centaurs from \citet[][H12]{Hainaut2012}, \citet[][F21]{Fernandez-Valenzuela2021}, and \citet[][B22]{Buchanan2022}. The colors of the 4-taxon system of \citet{Barucci2005} are shown by the rectangles \citet{Hasegawa2021}. Estimates of (B-R) of the \citet{Buchanan2022} measurements were done by determining the range of linear spectral slopes that match the reported (g-r) colors, and projecting those slopes onto (B-R). (R-J) was estimated from (r-J) in a similar fashion. }}
\label{fig:TNOtaxonomy}
\end{center}
\end{figure}

The Red/Very Red taxonomy only considers optical spectral slope or color, and does not make use of any other wavelengths. Longward of $\sim0.9 \mbox{ $\mu$m}$, TNOs tend to exhibit NIR colors that are correlated with optical spectral slope \citep[e.g.][]{Fraser2012}. Attempts have been made to use the optical-NIR color space to create more complex taxonomies. 
Early efforts applied principal component analysis to a sample of BVRIJ colors and found 4 classes \citep[e.g.][]{Barucci2005}. 
This 4-taxon system \textcolor{black}{is presented in Figure~\ref{fig:TNOtaxonomy},} and essentially divides the Red and Very Red classes into two subpopulations based on their IR behavior. For example, the IR and RR classes are Very Red, with the RR exhibiting red spectral slopes across the BVRIJ range, and the IR exhibiting bluer NIR spectral slopes than found in the optical. 
Generally, the number of taxa found would increase with the number of filters or the complexity of the analysis \citep{DalleOre2013}. 
A recent effort to include visible albedo in the analysis expands the number of taxa to as many as 6, with an additional 4 taxa each containing a single dwarf planet (Pluto, Eris, Makemake, and Quaoar). Alternative analyses have concluded that the optical-NIR color distribution should not be further subdivided. Rather, excited TNOs exhibit only 2 separate taxa, with each exhibiting a continuum of colors through the optical and NIR color spaces \citep{Fraser2012,Schwamb2019}. 

For completeness, we point out the so-called cold classical TNOs, which stand out in many ways, including their tight orbital distribution (see Section~\ref{sec:kuiperpops}) and high binary fraction \citep{Noll2008,Noll2020}. 
They are found in a tight annulus between $42<a<48$ with inclinations $i\lesssim5^o$. Considering their red optical colors \citep{Gulbis2006}, cold classicals mainly appear to belong to the Very Red taxon. 
Unlike most dynamically excited Very Red objects, however, cold classicals appear to exhibit NIR colors that tend to values closer to solar \citep{Pike2017} and have higher than usual visual albedos \citep{Brucker2009}, suggesting cold classical objects may occupy a third taxon. 
Due to their stable orbits, however, they do not act as a significant source of Centaurs or JFCs.

Unsurprisingly, the Centaurs tend to exhibit similar optical-NIR color distributions as the TNOs that feed the Centaur population. 
A detailed comparison is limited by biases, both observational, such as inconsistent target selection in color surveys \citep{Schwamb2019}, and physical, such as that colors seem to vary with size \citep{Benecchi2019}, and surface alteration processes \citep{Seccull2019}, all of which will likely affect the TNO and Centaur color distributions differently.

JFCs do not exhibit any of the color structure shown by TNOs and Centaurs. Instead, they consist entirely of objects with surfaces only slightly redder than solar \citep[e.g.][]{Jewitt2015}. 
As the JFCs are predominantly directly fed from the excited TNO populations, the gray surfaces of the JFCs must be the result of alterations to their surfaces as objects migrate closer to the Sun. Cometary activity seems a likely culprit. Indeed, those Centaurs which have been found to be active all have surface colors that are closer to solar than the average TNO, with none of the Centaurs with Very Red colors exhibiting detectable levels of activity. This implicates activity as a main driver for surface color alteration, possibly through deposition of gray dust \citep[e.g.][]{Seccull2018}. Other alteration mechanisms might be at work, such as the thermal destruction of reddening agents.

\subsection{Coma Compositions of Centaurs}
\label{sec:Centaur_comps}

At least 10--15\% of known Centaurs exhibit dust comae and are deemed active \citep{Jewitt2009, Bauer2013}. Such comae are composed primarily of dust grains with expansion velocities of 0.3 km s$^{-1}$ or less and production rates ranging from $\sim$ 1--1,000 kg s$^{-1}$ (Table \ref{tab:dustgas}) \citep{Whitney1955, Hughes1991,
Trigo-Rodriguez2010, Kulyk2016}. Identifying the volatile components and ascertaining which are produced in high enough amounts is an important first step in constraining models of Centaur activity and solar system formation and evolution. 

\textcolor{black} {A summary of volatile production rates in Centaurs is provided in Table \ref{tab:dustgas}. There are so few detections and significant limits that their production rates (in molecules s$^{-1}$ and kg s$^{-1}$) can be listed succinctly in this table. Representative dust mass loss rates are also included in the table, with 29P leading the pack with the highest values, although low rates are sometimes seen in this Centaur.}

\begin{table*}[!ht]
\centering
\caption{Measured and derived properties of dust and gas in Centaurs}  \label{tab:dustgas}
\begin{tabular}{|l l r r r r|}
 \hline
{\bf Comet} & {\bf Volatile} & {\bf r (au)} & {\bf Production rate} & {\bf Mass loss rate $^a$
} & {\bf Reference} \\
  & &  & 10$^{27}$ mol/s & (kg/s)  & \\
 \hline
29P & Dust  & 5.8 - 6.3 &  -- & 30 -- 4700  &   [1,2,3,4,5] \\
& CO & 5.7 - 6.3 & 10 - 70 & 460 -- 3200  &   [6,7,8,9,10,11]\\
& H$_2$O & 6.2-6.3 & 4.1 - 6.8  & 120 - 188   & [5, 8] \\
& CO$_2$ & 6.2 & 0.5 $\pm$ 0.3  &  36 $\pm$ 22   &  [12]\\
&  N$_2$ (N$_2^+$) $^b$ & 6.3  & 0.4 $\pm$ 0.2 & 17 $\pm$ 8  &  [10, 11, 13] \\
& O$_2$ $^c$ & 6.3 & $\sim$0.24 & $\sim$13 & [11] \\ 
& HCN  & 6.3 & 0.048 $\pm$ 0.011 & 1.8 $\pm$ 0.4 &  [5] \\
& CN & 5.8 & 0.008 $\pm$ 0.003  &   0.3 $\pm$ 0.1  & [14]   \\
& NH$_3$ & 6.3 & $<$4.5 & $<$128 & [9]  \\
& CH$_3$OH  & 6.3& $<$0.55 & $<$29 &  [9] \\
& CH$_4$  & 6.3& $<$1.3 & $<$34 &  [9]\\
& H$_2$CO & 6.3 & $<$0.1 &$<$5 &  [9] \\
\hline
Chiron & Dust & 9 - 18 & - & 1 - 45 & [15, 16, 17]\\
& CO & 8.5 & 13 $\pm$ 3 & 605 $\pm$ 151& [18]\\
& CN &  11.3 & 0.037 $\pm$ 0.011 & 1.4 $\pm$ 0.3  &  [19]\\
\hline
Echeclus & Dust & 13.1 & - & 10 - 700 & [20, 21]\\
&  CO& 6.1 & 0.8 $\pm$ 0.3 & 300 $\pm$ 130 & [22] \\
& CN & 12.9 & $<$0.04 & $<2$ &[23] \\
\hline
P/2019 LD$_2$ & Dust & 4.5 &- & 40 - 60& [24] \\
& CO & 4.58 & $<$4.4 & $<$200 &[25]  \\
& CN & 4.5 & $<$0.0014 &  $<$ 0.1 & [23] \\
\hline
\end{tabular}
\begin{flushleft}
[1] \citet{Fulle1992}, [2] \cite{Ivanova2011}, [3] \cite{hosek2013} [4] \cite{Schambeau2021} [5] \cite{BockeleeMorvan2022}, [6] \cite{SenayJewitt1994} [7] \cite{festou2001}, [8]\cite{Ootsubo2012}, [9] \cite{Paganini2013} [10] \citet{Womack2017},  [11]  \cite{Wierzchos2020}, [12] Harrington Pinto et al., under review, [13] \cite{Ivanova2016}, [14] \cite{Cochran1991CN}, [15] \cite{luujewitt1990}, [16] \cite{romonmartin2003}, [17] \cite{fornasier2013}, [18] \cite{WomackStern1999}, [19] \cite{Bus1991}, [20] \cite{Rousselot2016}, [21] \cite{Bauer2008}, [22] \cite{Wierzchos2017}, [23] \cite{rousselot2008}, [24] \cite{Licandro2021}, [25] \cite{Kareta2021}, 
\\
\footnotesize{$^a$ Gaseous mass loss rates were calculated using production rates and appropriate atomic mass units.}\\
\footnotesize{$^b$ N$_2$ production rate is derived from CO$^+$, N$_2^+$ and CO measurements, see \citet{Womack2017}.}\\
\footnotesize{$^c$ O$_2$ production rate is inferred by scaling the mixing ratio of O$_2$ and H$_2$O in 67P (a JFC) to the water production rate measured in 29P by AKARI, see \citet{Wierzchos2020} for details.}\\
\end{flushleft}
\end{table*}

\textcolor{black}{Despite decades of observations, very few volatiles have been detected on Centaurs. This is largely due to their faintness and because most are inactive. Their distant orbits combined with only rare episodes of comae or outbursts  make it difficult to discover when they are active and then carry out observations. Challenges also arise from the fact that some of the candidate species for driving activity, such as CO$_2$, CH$_4$, N$_2$, and O$_2$, are rotationally symmetric and thus have no pure rotational transitions at mm-wavelengths, which is how most cometary volatiles are observed. Furthermore, telluric contamination is a substantial obstacle for detecting CO$_2$ and CH$_4$ emission via infrared rovibrational spectra, and thus observing these two volatiles directly has been restricted to a few instruments above Earth's atmosphere, such as the Infrared Space Observatory, AKARI Space Observatory, and in situ spacecraft mission measurements.} 

Highly volatile species, such as N$_2$ and O$_2$, may be abundant enough in nuclei to play a role in Centaur activity, but thus far have not been detected. \textcolor{black}{N$_2^+$, a probable ionization product of N$_2$, has been detected in optical spectra of 29P's coma. N$_2^+$ is also a frequent night sky emission feature; however, a telluric source can be ruled out with longslit spectra when N$_2^+$ emission is visible only in the tailward direction (e.g., \citet{Ivanova2019}). Analysis of 29P's N$_2^+$ and CO$^+$ emission in such tailward spectra of 29P is consistent with an N$_2$ production rate of $\sim$ 17 kg s$^{-1}$, which is much smaller than CO \citep{Wierzchos2020}. However, care should be used when interpreting either N$_2^+$ and CO$^+$ spectra, because their line strengths in 29P's coma appear strongly correlated with solar wind particle velocities, possibly indicating a charging mechanism of solar-wind proton impact onto CO and N$_2$ in the coma  \citep{Cochran1991, Ivanova2019}.}  An O$_2$ production rate was estimated by using measurements of its abundance in the Rosina mass spectroscopy of the JFC 67P, and then scaling it to the measured water production rate \citep{Wierzchos2020}. These values suggest that N$_2$ and O$_2$ are not produced in high enough amounts to drive much of the activity of 29P. 

HCN emission was detected at millimeter wavelengths in only one Centaur -- 29P in 2010 -- but with relatively low production rates and with a spatial profile that indicates it may have been released from icy grains and not directly sublimating from the nucleus \citep{BockeleeMorvan2022}. Although seen in TNO and Centaur reflectance spectra, methanol (CH$_3$OH) emission has not yet been detected in Centaur comae, with the strongest upper limit of 29 kg s$^{-1}$ set for 29P \citep{Biver1997}. 

Evidence for water-ice grains has been observed in cometary comae \citep{Yang2009, A'Hearn2011, Kelley2013, Protopapa2018}; by extension, it is reasonable to search for them in Centaur comae. Centaurs orbit at heliocentric distances too large for water ice on the nucleus surface to sublimate efficiently, but water vapor may still be detected in the coma due to water ice-rich grains that are carried off the nucleus by other outgassing mechanisms and then sublimate once bathed in the solar radiation field, as was reported for 29P for water  \citep{Ootsubo2012} and in the aforementioned HCN. Recently, the signature of water ice grains was again reported during a significant outburst of 29P in September 2021 \citep{Kelley2021}, and a water ice grain signature at the 1--10\% level was seen in the near-infrared spectrum of the Centaur C/2019 LD$_2$ (Pan-STARRS)'s coma at 4.6 au \citep{Kareta2021}. 

There are also occasional reports of CN, C$_2$, and C$_3$ detections or upper limits. These are radicals and not capable of long-term storage in the nucleus, and so are probably daughter products from another cometary volatile and not directly responsible for driving activity \citep{Bus1991, Cochran1991, Womack2017, Ivanova2019, Kareta2021, Licandro2021, Bolin2021}. 

Although considered as a likely candidate for distant activity in Centaurs because of its relatively high abundance in many comets and the good match of CO$_2$'s sublimation efficiency at Centaur distances, CO$_2$ emission has been searched for in Centaur comae, with no detections thus far. A significant limit was obtained only for 29P with the AKARI space telescope \citep{Ootsubo2012}, implying a CO/CO$_2$ ratio $> 90$, much higher than what is seen in other comets (Harrington Pinto et al., under review). The space-based telescopes Spitzer and NEOWISE have been used to observe many Centaurs and were set up to detect the combined emission of CO and CO$_2$  in the same $\sim$ 4.5~$\mu$m filter bandpass. \textcolor{black} {Unfortunately, because the emission is combined, individual CO$_2$ or CO production rates cannot be derived from the data without independent and simultaneous measurements of one of the molecules.} \textcolor{black} {Recently, CO$_2$ production rates were inferred for 29P  using Spitzer and NEOWISE imaging data in 2010 along with contemporaneously obtained CO millimeter-wavelength spectra, which confirms that CO$_2$ is produced in relatively small amounts in the 29P coma, as low as approximately 1\% of CO and 14\% of H$_2$O, (Harrington Pinto et al., under review).}

Consequently, CO is the most tractable volatile for direct measurement in Centaur comae that may play a significant role in activity, and these observations are possible using ground-based as well as space-based instruments at millimeter and infrared wavelengths. Detection of CO emission is reported for three Centaurs: 29P, 95P/Chiron, and 174P/Echeclus \citep{SenayJewitt1994,WomackStern1999,Wierzchos2017, BockeleeMorvan2022}. On a related note, CO emission is also detected in some long-period comets at ``Centaur distances" from the Sun, such as C/1995 O1 (Hale-Bopp) and C/2017 K2 (Pan-STARRS) \citep{Gunnarsson2003,Yang2021} and CO appears to play a large role in distant comet activity \citep{Biver2002, Womack2017}.  

The apparent absence, or at least very low abundance in the coma, of CO$_2$ emission in 29P's coma is instructive. If we assume that Centaurs and comets formed in similar environments and should have similar chemical compositions of their nuclei, then the available data on comets may be useful for predicting what we should see in other Centaurs. In a study of 25 comets for which CO and CO$_2$ were simultaneously measured, the mixing ratio of CO and CO$_2$ production rates showed a possible preference for CO$_2$ in the comae of most JFCs, but the opposite is true for some Oort Cloud comets and the lone Centaur (29P),   \citep[see][and Harrington Pinto et al., under review]{A'Hearn2012} . However, if JFCs, active Centaurs, and long-period comets are analyzed as an aggregate, the coma mixing ratio may be better explained as following a heliocentric trend out to at least 6 au. CO$_2$ is preferentially detected in most comets (not just JFCs) within 3 au, and CO is dominant beyond 3 au, although selection effects cannot be ruled out since JFCs are typically fainter than long-period comets and less likely to be seen at large distances. However, if this trend continues for other active Centaurs, then CO emission is likely to exceed CO$_2$ emission in the comae of active Centaurs, as is the case for 29P. This effect may be at least partly due to differences in the degree of thermal processing of the nucleus, rather than compositional differences \citep{A'Hearn2012, HarringtonPinto2019}. Indeed, one of the problems remaining in cometary science is accurately determining comet nucleus composition from coma abundances.

\section{Activity and Physical Evolution: Comet Beginnings}

\subsection{Activity in Centaurs}
\label{sec:centaur-activity}

The drivers of activity in Centaurs have implications for the nature and evolution of ice in small bodies and the cause of all activity in icy bodies at large heliocentric distances. Unfortunately, very little is known about the dominant mechanisms. 
Active Centaurs frequently display two types of activity: sustained comae and discrete outbursts. The first Centaur discovered, 95P/Chiron,  undergoes periods of sustained activity as well as short-term brightening periods. Arguably, the \textcolor{black}{best-known} active Centaur is 29P, an object on a nearly circular orbit just beyond Jupiter's distance that exhibits continuous activity with occasional outbursts of 10--250 times in brightness superimposed. Another well known active Centaur is 174P/Echeclus, which has undergone several outbursts, including a large one in which the activity was centered on an apparent fragment of the main nucleus, and a few longer-term periods of activity with lower-level dust comae.

\textcolor{black} {Vigorous water-ice sublimation in comets releases many minor species in cometary comae, but Centaurs are too far from the Sun for water ice to sublimate efficiently. Thermal equilibrium temperatures of Centaur nuclei are most similar to the sublimation temperatures of CO$_2$ and NH$_3$ and hence they are favored by some models of Centaur activity, e.g. for the Centaur C/2014 OG$_{392}$ at $\sim$ 10 au \citep{Chandler2020}. However, no emission from these volatiles has yet been detected in a Centaur, and studies of 29P place stringent limits on their mixing ratio relative to CO.} 

As discussed in the previous section, CO is the only molecule detected in the gaseous state in active Centaurs that is capable of driving activity.  CO is also abundant in a few Oort Cloud comets active beyond 5 au, but no CO detections or significant limits exist for JFCs this far out (this may not imply compositional differences between JFCs and Centaurs). There is evidence for water ice grains in the comae for two Centaurs: 29P and P/2019 LD$_2$ (ATLAS). These grains were most likely released from a surface component, similar to what has been reported for some Centaur and TNO surfaces in Section~\ref{sec:Centaur_comps}.

Another model for Centaur activity invokes cosmogonically important volatiles with lower sublimation temperatures, such as CO, which can survive in the nucleus as ices just below the surface, and also be partially incorporated in the gaseous state within amorphous ice and then released when this ice undergoes the phase change to the crystalline state \citep{Prialnik1995, DeSanctis2000, Prialnik2008, Capria2009, Guilbert-Lepoutre2011}.  This phase change is optimized around $\sim$ 120-140K (equivalent to Centaurs and comets at 5--10 au), but it can proceed at farther distances (and colder temperatures) less efficiently.

The most popular model that explains the observations of Centaur (and distant comet) comae is that of a nucleus with pockets of amorphous ice that undergoes the crystallization process, which releases trapped volatiles. Another model invokes isolated and insulated pockets of frozen CO or CO$_2$ that sublimates once the surface is disrupted enough to expose a relatively small and previously protected frozen patch \citep[e.g.][]{Prialnik2008}. However, given the very few measurements we have on Centaurs, more observations are needed to settle between these two models. 

\begin{figure}[htpb]
    \centering
    \includegraphics[width=0.49\textwidth]{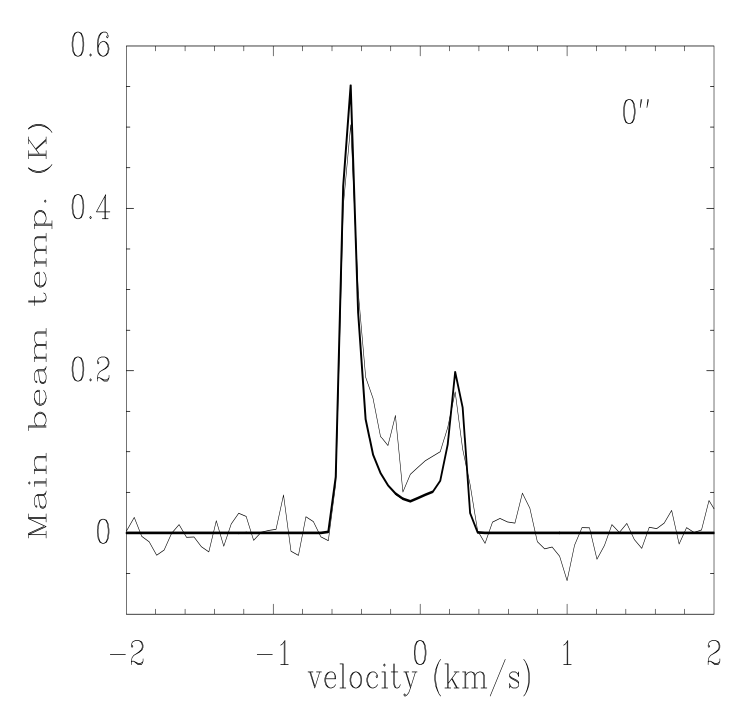}
    \caption{Spectral line of CO J=2-1 rotational transition at 230 GHz for the Centaur 29P/Schwassmann-Wachmann 1 obtained with the IRAM 30-m telescope in January 2004. Superimposed is a model fit to the data, which includes a strong blue-shifted peak of sunward emission from a very cold nucleus source, and a smaller contribution from the nightside. From \citet{Gunnarsson2008}}.
    \label{fig:COCentaur}
\end{figure}

Because of the lack of detections or low production rates and significant limits for  other volatiles in Centaurs, our main clues to outgassing must come from CO in 29P. The spectral line profile of CO via its J=2-1 rotational transition at 230 GHz is  very narrow and slightly blueshifted, consistent with sunward emission of CO from a very cold region and a secondary source more distributed over the nucleus surface (Figure~\ref{fig:COCentaur}). This is also true for CO in Chiron \citep{WomackStern1999} and Echeclus \citep{Wierzchos2017}, as well as other distantly active comets like Hale-Bopp \citep{Biver1997, womack1997} and C/2017 K2 (Pan-STARRS) \citep{Yang2021}. The strikingly similar spectral shapes suggest there is a common outgassing mechanism for CO in active Centaurs and distant Oort Cloud comets  \citep{WomackStern1999, Biver2002, Gunnarsson2003, Gunnarsson2008, Wierzchos2017, Yang2021} originating from a very cold ($\sim$ 4K) gas, \citep[e.g.,][]{Paganini2013}. Thus, it is reasonable to plan for narrow emission from a very cold gas in the inner coma when searching for CO in other active Centaurs. 

\begin{figure}[htpb]
    \centering
    \includegraphics[width=0.49\textwidth]{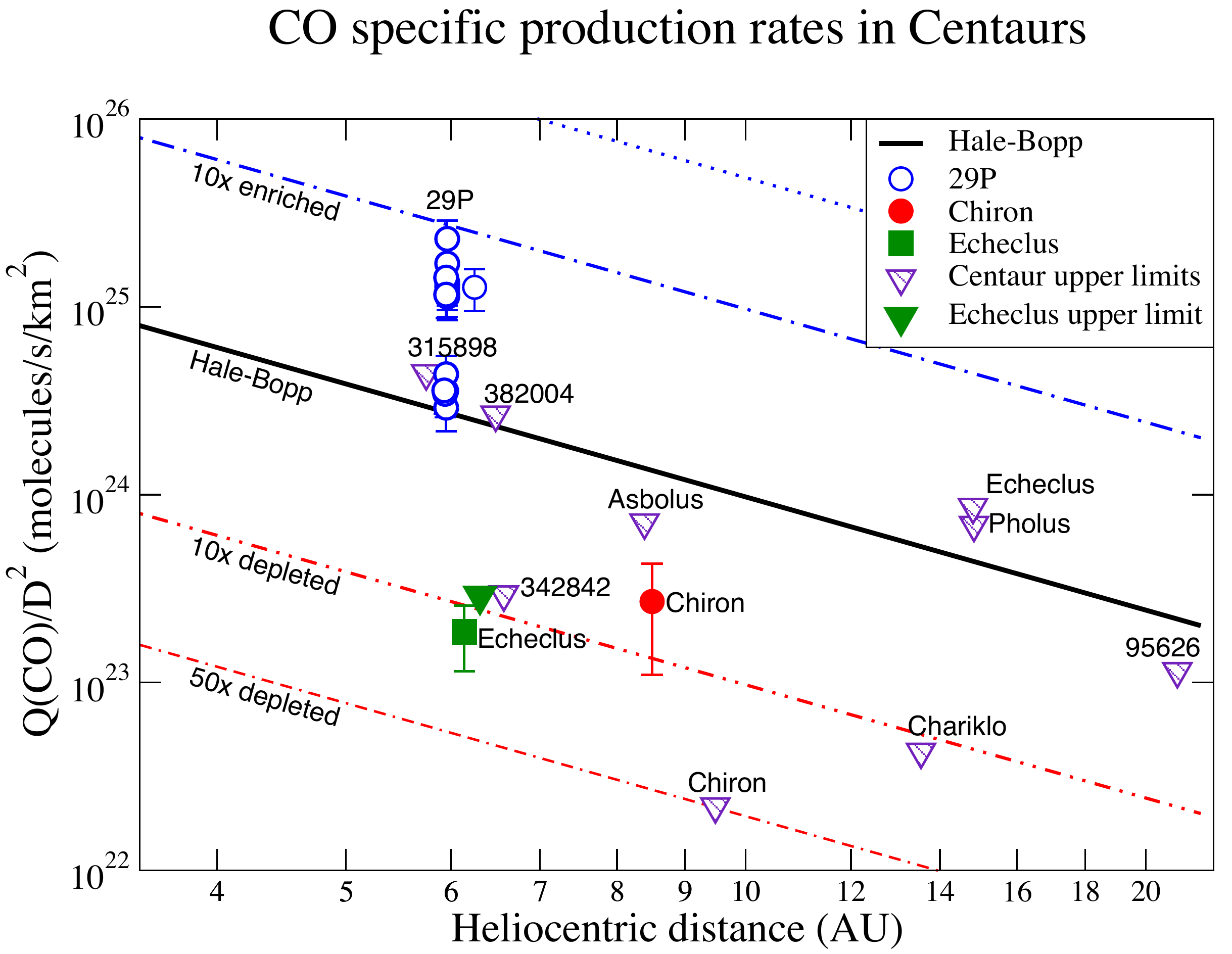}
    \caption{Specific gas production rates, Q(CO)/D$^2$, for Centaurs. The solid line is a fit to data for Hale–Bopp provided for comparison. As the figure shows, after normalized by surface area, 29P produces far more CO than other Centaurs like Echeclus and Chiron, where CO was detected in lower amounts, and other Centaurs for which strong upper limits were set. From \citet{Wierzchos2017}.
    }
    \label{fig:CentaurCO}
\end{figure}

Interestingly, Hale-Bopp, 29P, and Echeclus appear to have approximately the same nucleus diameter and were all observed at 6 au, making an intercomparison possible that minimizes size and heliocentric distance contributions \citep{Wierzchos2017}. Thus, 29P is an abundant producer of CO, even sometimes outproducing Oort Cloud comet Hale-Bopp at the same distance. In contrast, Echeclus and Chiron emitted CO very weakly, and it was only marginally detected. Searches for CO in many other Centaurs did not yield detections \citep{Drahus2017}, but some, such as (10199) Chariklo, 342842 (2008 YB3), (8405) Asbolus, and 95626 (2002 GZ32) provide tight upper limits that eliminate the possibility of significant outgassing activity from CO at the time of observation (see Figure~\ref{fig:CentaurCO}).  We do not know whether Centaurs like Chiron, Echeclus, and others that produce little to no CO formed in a different  environment, or whether they are devolatilized, or have not yet started to become more active, partly due to spending more of their orbits well beyond 6 au, when compared to 29P. The difference in CO output in 29P and these other active Centaurs is particularly striking and may be useful to constrain Centaur models. Further accurate measurements of nucleus diameters and CO and CO$_2$ production rates are needed to better understand how these species contribute to activity and ultimately to determine the composition of their nuclei.

Centaur activity also appears to be related to the residence time in its orbital region, and a decrease in perihelion distance and/or semimajor axis 
may occur before the observed onset of activity \citep{Fernandez2018, Sarid2019, Lilly2021}. A Centaur's perihelion distance also apparently plays a large role in whether a Centaur will become active, with lower values more likely to correlate with activity \citep{Jewitt2009}. \citet{Rickman1991} noted a similar trend for short-period comets whose perihelion distances had recently decreased. Such comets had larger values for their nongravitational parameters, suggesting that their mantles had been removed, allowing more of their surfaces to be active. 
\textcolor{black} {The ongoing CO outgassing in Centaur 29P is recently proposed to be due to the nucleus responding via the crystallization phase change of water due to the relatively sudden, within $\approx 2000$~years, the median residence time of the Gateway orbit \citep{Sarid2019}), change in its external thermal environment produced by its dynamical migration from the Kuiper belt to the Gateway region where it maintains a nearly constant thermal environment at a heliocentric distance of $\approx6$~au (Lisse et al. 2022, under review).}
Thus, orbital history may play a large role in triggering the activity of Centaurs.

In addition to steady-state production that may produce a coma lasting weeks or months, active Centaurs sometimes have discrete outbursts lasting a few days to weeks. 29P is the best-known example whose outbursts have been documented for more than a century, primarily with secular lightcurves of the visible magnitudes and images which capture morphology in the coma, e.g., \citet{Trigo-Rodriguez2010, Schambeau2017, Schambeau2019}. One of the largest outbursts in decades began in September 2021, when 29P increased its brightness by a factor of $\sim$ 250 during a series of four smaller outbursts occurring over a few days. Such outbursts give an opportunity to test models of chemical composition and physical mechanisms of ejection. Unlike 29P, Chiron and Echeclus typically do not maintain long-term quiescent comae, but both have exhibited outbursts or brightening episodes. In 2005 Echeclus underwent a $\sim$ 7 magnitude outburst (from 21 to 14, corresponding to an increase in brightness by a factor of $\sim$ 630) and was accompanied by a large, detached coma $\sim$ 2 arcmin across, projected to be 1,000,000 km at the Centaur's distance and was visible for a few weeks \citep{choiweissman2006, tegler2006, rousselot2008}. More details about its observational timeline, orbital dynamics and measured characteristics during outbursts are found in \citet{Wierzchos2017, Seccull2019, Kareta2019}. 

29P's long-term behavior of continuously having a dust and gas coma, regularly punctuated by outbursts, is well-suited to dedicated observation over years, and its lightcurve displays two types of outburst shapes: most have an asymmetric sawtooth pattern,  where the reflected light grows tremendously in a few hours and then decays in days to weeks. These outbursts must have an explosive trigger that releases a great deal of dust. Less frequently, the visible lightcurves of smaller outbursts appear to grow and decay symmetrically in time  \citep[cf.][]{Trigo-Rodriguez2010, Wierzchos2020, Clements2021}.  In contrast to the larger sawtooth shaped outbursts, the smaller symmetric outbursts may originate from a source region that spans a larger surface area on the nucleus which releases material over a longer period of time, perhaps a few days. \textcolor{black}{Still another mechanism contributing to the changing brightness of Chiron is the presence of rings or ring arcs (also detected around Chariklo \citep{Braga-Ribas2014}),
which changes the Centaur's apparent brightness as the aspect of the rings varies from open to edge-on \citep{Sickafoose2020, Fernandez-Valenzuela2022}.}

There are very little data about possible triggers for the outbursts in 29P. However, outbursts of CO and dust were both recorded during simultaneous observations of CO mm-wavelength spectral flux measurements and visible magnitudes, providing an opportunity to test the hypothesis that a strong CO outburst was needed to trigger the dust outbursts. Interestingly, the CO production rate doubled, but did not trigger a noticeable rise in dust production \citep{Wierzchos2020}. Similarly, two dust outbursts occurred without an accompanying increase in CO production. Two other dust outbursts may show CO gas involvement. These odd results may be explained if the CO is not always substantially incorporated with the dust component in the nucleus, or if CO is primarily released through a porous material. 

\subsection{Morphological Evolution and TNO Binaries in Context}
\label{sec:morphology}

\textcolor{black} {The prevalence of contact binaries found in the comet and TNO populations is suggestive of a formation mechanism that preferentially forms contact binary objects. Such a connection is not so obvious, however, without morphological and evolutionary considerations.} 
This short discussion highlights only a few of the important processes
one needs to consider when comparing the morphologies of JFCs to their
precursor TNO populations, and the challenges faced when attempting to
make inferences about their formation modes.

Most cometary nuclei imaged by spacecraft or radar have bilobed shapes; of the seven, only 9P/Tempel 1 and 81P Wild are nearly spheroidal 
(see the chapters by Pajola et al.\ and Knight et al.\ for further details). 
This suggests that comets are born bilobed or
morphologically evolve to become bilobed during their lifetimes.  Here
we first consider the formation stage.

Formation of small planetesimals in the outer solar system favors
binarity.  For example, the streaming instability model \citep[][see chapter by Simon et al.]{Youdin2005}, where rotating clouds of
pebbles collapse under their own gravity, gives birth to 
binary systems with near--equal-sized components, with properties that match observations \citep{Grundy2019,Nesvorny2019-binNatAst}.
 The components of a newly formed close binary
can be brought into contact by gas drag \citep{Lyra2019}, producing
a contact binary. The early-stage low-speed collisions between similar
size bodies can lead to mergers and bilobate shapes as well \citep{Jutzi2015}.

Contact binaries are ubiquitous in the Kuiper Belt. For
example, the New Horizons spacecraft revealed that
Arrokoth is a contact binary, which likely resulted from a low-speed merger
of two flattened, spheroidal components \citep{Stern2019}. Additionally, light
curve observations of TNOs indicate that the contact binary fraction
can be 30\% or higher \textcolor{black} { \citep{Sheppard2004, Thirouin2018, Thirouin2019, Thirouin2022, Noll2020, Showalter2021}.}

It is suggestive to draw connections between these observations and
bilobed comets. Note, however, that: (i) comets are much smaller than
most known TNOs, and (ii) Arrokoth, with its nearly-circular,
nearly-ecliptic orbit 44 au from the Sun, formed well beyond the
original formation region of most comets, in a sparsely populated
region of the Kuiper Belt. 
As discussed in Section~\ref{sec:origins_tnos}, present-day comets presumably formed in a massive disk $\sim$20--30 au from the Sun, became part of the scattering TNO population (at distances of $\sim$~50--1,000 au) for $\sim4.5$~Gyr, then evolved onto inner solar system orbits.
The size and formation distance of TNOs affect their survival. Arrokoth is thought
to be a pristine planetesimal that formed and survived essentially unchanged in the low-mass classical belt
\citep{McKinnon2020}. 
Small comets, instead, are less likely to
survive intact, their first obstacle being the disruptive and
shape-changing collisions during the massive disk stage \citep{Benavidez2022}.

The overall significance of collisions during the disk stage mainly depends on the disk lifetime, $t_{\rm disk}$. 
If $t_{\rm disk} \gtrsim 10$ Myr, the great majority of comet-sized bodies would be disrupted \citep{Morbidelli2015}, and the size distribution of small bodies would approach the Dohnanyi slope \citep{Dohnanyi1969,O'Brien2003}.  
This could explain the size distribution break observed near $D=100$ km \citep[][ and Section 4.3]{Bernstein2004, Bernstein2006, Fraser2014}. Any traces of the
original surface morphology would be wiped out: most comets would be
fragments of larger bodies. If $t_{\rm disk} \ll 10$~Myr instead, 67P-sized
comets would avoid being catastrophically disrupted. Smaller impacts,
however, could still cause important shape changes
\citep{Jutzi2017}.

Catastrophic disruptions \citep{Schwartz2018} 
and sub-catastrophic
impacts on elongated and rotating bodies have been modeled to
demonstrate the formation of bilobate comets \citep{Jutzi2017}. As the
massive disk was dispersed by Neptune, the collisional probabilities
dropped, and the collision speeds increased from hundreds of m/s to
several km/s \citep{Nesvorny2019}. It is likely in this
situation that each catastrophic disruption was followed by
sub-catastrophic impacts, and each sub-catastrophic impact --
potentially capable of generating a bilobate shape -- was followed by
shape-changing impacts. The observed comet shapes could be a complex
end product of this sequence.

As comets evolve into the inner solar system, they become affected by
H$_2$O sublimation torques. Simulations suggest that 67P, and bilobed
comets in general, should often spin up past the breakup limit,
fission, and reconfigure \citep{Hirabayashi2016}.  Many new bilobate
configurations can be produced by this process. CO or CO$_2$
sublimation-driven spin-up might be capable of disrupting
a typical JFC even before it reaches the inner solar system \citep{Safrit2021}.  These results highlight the important relationship between
spin and morphology. Adding to that, a spinning bilobed object also
better resists reconfiguration into a more spheroidal body by small
impacts (Jutzi et al. 2017).

There are at least two basic possibilities: (1) small TNOs formed as contact binaries and these shapes survived, even if with some modifications, such that TNOs remained bilobed as they dynamically evolve into Centaurs/JFCs; or (2) the bilobed shapes of JFCs have nothing to do with the formation of TNOs, but instead were produced by other processes, such as impact disruption, activity \citep{Safrit2021}, rotational fission, etc., and could reflect the reaccumulation of fragments. It may be hard to distinguish between these two possibilities. One option would be to measure the shapes of a large sample of small TNOs via occultations or lightcurve surveys, and, in the long run, via spacecraft imaging. This would give us a rough sense of the fraction of contact binaries in different TNO populations, including the cold classicals. For example, option (1) could be ruled out if small cold classicals show a very low contact binary fraction. This would indicate that small planetesimals in the outer solar system did not form bilobed, therefore pointing toward evolutionary processes as the primary cause.  It is worth pointing out, however, that the only small TNO we have flown past happens to be bilobed. We could also expect some information from future surveys about the physical properties of TNO contact  binaries -- for example, about the relative size of individual components in each detected small contact binary  -- and compare that with bilobed JFCs and Centaurs. A match would be expected if small TNOs formed bilobed and the evolutionary processes have only a small effect on the overall shape.

\subsection{Size Distributions of TNOs, Centaurs, and JFCs}\label{sec:sfds}

Here we consider the size-frequency distribution (SFD) of TNOs, Centaurs, and JFCs. The SFD holds information about the formative, morphological, and destructive processes that have altered the sizes of these bodies. 

The SFD of Trans-Neptunian Objects is almost always estimated from reflectance photometry of TNOs in well-characterized surveys. 
The cumulative distribution of apparent magnitudes $m$, which we will call the luminosity function (LF), is typically expressed as $\Sigma(<\!m) \propto 10^{\alpha m}$. 
Here $\Sigma(<\!m)$ represents the surface density in TNOs per square degree on the sky (usually on the ecliptic) of bodies with magnitudes less than (brighter than) $m$, and $\alpha$ is the slope of the distribution on a plot of $\log_{10}{\Sigma}$ vs. $m$. 
If the TNOs follow a power-law differential size distribution of the form $dN/dD \propto D^{-q}$, where $D$ is the diameter of the TNO, and we assume constant albedos, then the \emph{differential} SFD slope, $q$, and  $\alpha$ are related by $q = 5\alpha + 1$. 
The {\em cumulative} size distribution is given by \textcolor{black}{$N(>D) \propto D^{-\gamma}$, where $\gamma = 5 \alpha = q - 1$ (e.g., \citet{Gladman2001})}. 
Correctly converting from a magnitude distribution to an SFD relies on a number of assumptions, including (1) the albedo and size distributions are constant and do not depend on heliocentric distance, (2) shape/lightcurve effects are not important, and (3)
observational biases such as the limiting magnitude of the survey are accounted for. 
With these limitations in mind, surveys revealed a steep LF in the outer solar system for objects brighter than $m_r\sim25$, with slope $\alpha\sim0.7$ or $q\sim4.5$
\textcolor{black}{\citep{Jewitt1998,Gladman1998,Trujillo2001,Fraser2008}}, significantly steeper than the value of $q$ between 3 and 4 expected for a population in steady-state collisional cascade \citep{Dohnanyi1969, Matthews2014,O'Brien2003}. 
Deep ``pencil-beam'' surveys revealed a shallower LF with slope $\alpha\sim0.2$ at fainter magnitudes, with a transition between the bright and faint slopes at a brightness $m_R\sim26$ \citep{Bernstein2004, Bernstein2006, Fuentes2009,Fraser2009}. 
This transition is referred to as the ``knee'' or break magnitude in the LF.

As surveys and observational techniques improved, more of the discovered objects were tracked over time, allowing accurate distance measurements and even dynamical classification of the survey discoveries.
We highlight the Canada-France-Ecliptic Plane Survey, which arguably was the first survey to provide 100\% tracking for all survey discoveries \citep{Petit2011}. 
This improvement enabled the first direct measurements of the \textcolor{black}{differential} absolute luminosity function (ALF), $\Sigma\left(H\right)$, where $H=m-2.5\log(\Delta r)-f(\alpha_p)$. Here \textcolor{black}{$m$ is the apparent magnitude of a TNO (often in $R$ band),} $r$ and $\Delta$ are the TNO's heliocentric and geocentric distances,  $\alpha_p$ is the phase angle, or observer-Sun-TNO angle, in degrees, and $f$ is a function that describes the decrease in reflectance of a TNO with increasing $\alpha_p$. 
For the low phase angles for TNOs accessible from the Earth, $f$ is usually linearly approximated as $f=0.15~\alpha_p$\textcolor{black}{, where $\alpha_p$ is measured in degrees }\citep[see][for further details]{Alvarez-Candal2016}. 
The ability to measure the ALF provided higher fidelity towards inferring the true underlying SFD.

The shape of the ALF has been characterized by power-law slopes of the form $\Sigma(H)=10^{\alpha_H(H-H_o)}$, and is known to exhibit three distinct regions. This functional form has been chosen as a matter of convenience, as it not only trivially translates to the power-law SFD discussed above, but also provides a statistically sufficient description of the observations \citep{Fraser2014}. Formative and collisional processes do not necessarily favor the production of power-law SFDs \citep[][see chapter by Simon et al.]{Li2019}.

The slopes of the ALF of the dynamically excited populations of TNOs are well described by

\begin{equation}
\alpha_H \sim
\begin{cases}
0.2, \text{ for } H_r\lesssim4\\
0.87,\text{ for } 4\lesssim H_r \lesssim 8\\
0.2,\text{ for } H_r>8.
 \end{cases}
\end{equation}

\begin{figure}[ht!]
    \centering
    \includegraphics[width=0.49\textwidth]{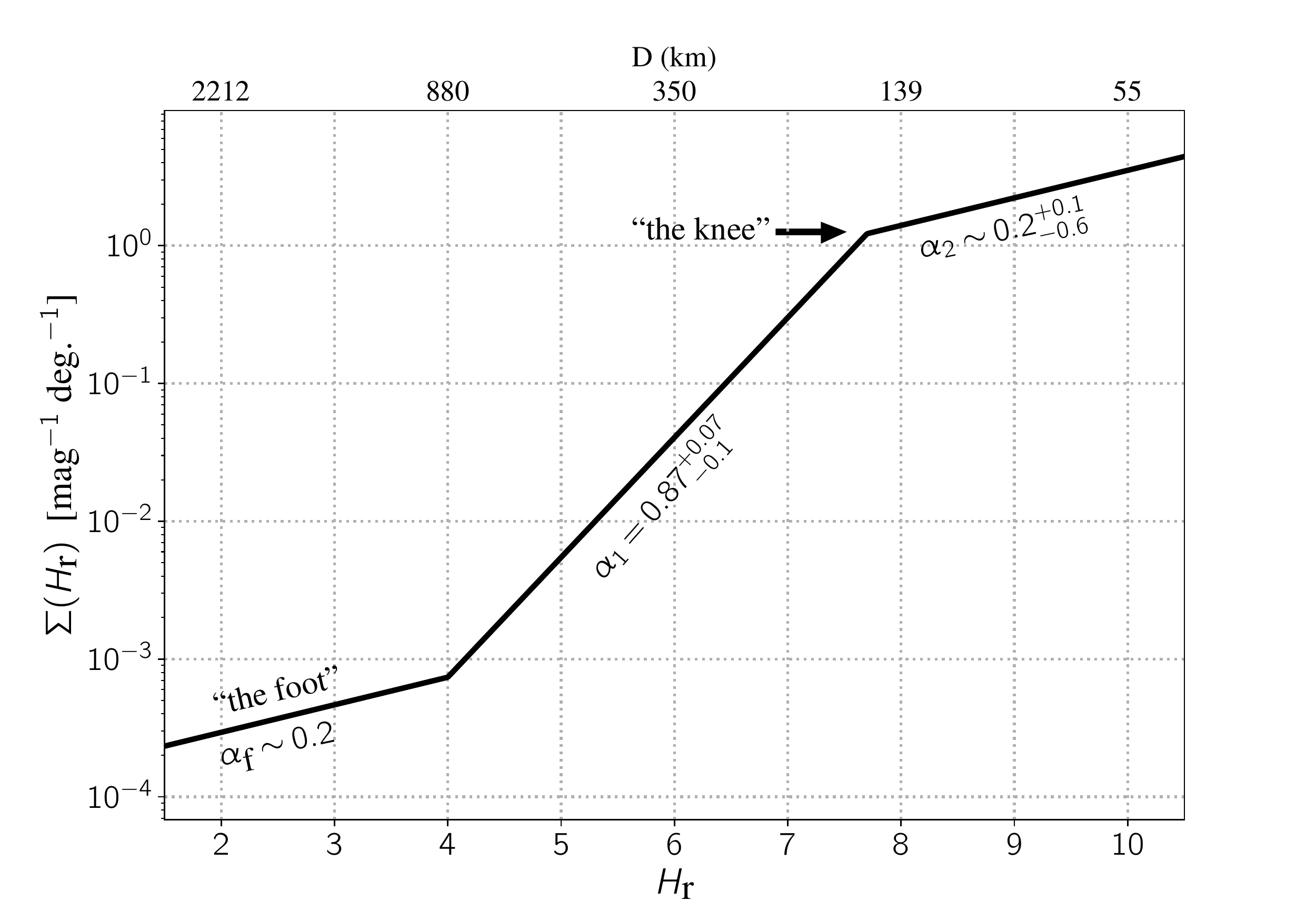}
    \caption{\textcolor{black}{A depiction of the measured differential absolute luminosity function of the dynamically excited TNOs. Each section of the ALF is approximated by the functional form $\Sigma(H) \propto 10^{\alpha(H-Ho)}$. Slopes and transition magnitudes are taken from \citet{Fraser2014}. Object diameters are shown across the top, and assume a geometric albedo of 6\%.}}
    \label{fig:sfd}
\end{figure}

\noindent
\textcolor{black}{We show a depiction of this shape in Figure~\ref{fig:sfd}.
This ALF translates to a size distribution that is shallow for the largest TNOs ($500\lesssim D\lesssim 2100$~km), with SFD slope $q\sim2$ for $D\gtrsim800$~km for a 6\% albedo \citep{Brown2008, Fraser2014, Nesvorny2017, Abedin2022}. }
This part of the size distribution is colloquially referred to as the \emph{foot} of the SFD \citep{Fraser2012dps}. 
The SFD has a steep slope $q\sim5.25$ for objects with $100\lesssim D \lesssim 600$~km and then becomes shallow again, with $q\sim2$, for sizes smaller than the knee, $D\lesssim100$~km \citep{Fraser2014}.
Direct survey constraints on the SFD much below the knee (e.g., to sizes $D\ll100$~km) are difficult to gather due to the faintness of such small objects in reflected light.

Two separate scenarios for the creation of the SFD of the hot population have been postulated. Historically, the break at $H_r \approx 8$ has been associated with the size above which catastrophic collisions rarely occur. That is, since the epoch of formation, collisions have been largely disruptive due to the high relative velocities between TNOs \citep[e.g.,][]{Dell'oro2013}. Also, for most of the solar system's planetesimal populations, including TNOs, smaller bodies are more numerous, and also easier to disrupt \textcolor{black}{down to sub-km sizes \citep{Benz1999}}. 
The net result is that smaller bodies are more likely to be collisionally disrupted than are relatively larger bodies \citep{Bottke2005}. It follows that most of the largest bodies may have avoided disruption. 
The steep slope for large objects then is a result of their formation process, originally thought to occur by hierarchical accretion, and the break diameter reflects the size below which, on average, all objects have been collisionally disrupted. Numerical simulations bear out this idea \citep[e.g.,][]{Kenyon2008,Benavidez2009}, 
though recently, this interpretation has fallen out of favor. It is now thought \textcolor{black}{that the observed break diameter is primordial, not collisional, in origin. If the break were collisional, the significant population of binary systems seen in the Kuiper Belt could not have survived \citep{Parker2010,Nesvorny2019}.}

The alternative idea to the collisional disruption scenario is the so-called born-big scenario, first put forth to explain the size distribution of the main-belt asteroids \citep{Morbidelli2009bornbig}, which has also been invoked to explain the break in the SFD of the Neptune Trojans \citep{Sheppard2010}. 
In this scenario, planetesimals form with a preferred size, $D\sim200$~km, in a formation process that is much more rapid than can be achieved through hierarchical accretion. One possible mechanism rests with the streaming instability discussed in Section~\ref{sec:morphology}. 
In this scenario, the break diameter reflects the preferred formation size, and the steep SFD at larger sizes reflects subsequent growth through classical hierarchical accretion. Notably, this scenario is broadly compatible with the observed properties of TNO binaries, including the frequency \citep{Nesvorny2011-binaries,Robinson2020}, orbital distribution \citep{Grundy2019-binaries,Nesvorny2021}, and diameters of TNOs themselves \citep{Li2019}.

In either scenario, the \emph{foot} of the SFD remains unexplained. This structure seems similar to expectations from models of runaway growth \citep{Lithwick2014}, though that has not been confirmed. It may also be that the foot is merely a signature of the largest objects that can form through the streaming instability. 

Extension of the SFD to smaller diameters comes from interpretation of the Pluto, Charon, and Arrokoth cratering records. \textcolor{black}{At diameters $D\gtrsim$~1--2~km, the data are consistent with} a slope $q \approx 3$ from the knee down to these sizes. 
For $D\lesssim$~1--2~km, the SFD appears to break again at the ``elbow'' to an even shallower slope, $q=1.7\pm0.3$, betraying a relative dearth of small TNOs with $0.1<D<1$~km, compared with an extrapolation from larger bodies \textcolor{black}{\citep{Greenstreet2015, Greenstreet2016, Singer2019, Parker2021, Robbins2021, Singer2021}}. A robust explanation of the change in the slope of the SFD near 1~km remains unavailable.

There is some hint of a so-called \emph{divot} in the size distribution, just smaller than the knee diameter. 
The divot is a purported sudden downward deviation in the SFD, with fewer objects at sizes just below the break than above it \citep{Shankman2013,Lawler2018}. 
Such a feature can result from a population of planetesimals that already has an SFD reminiscent of that observed (a steep slope followed by a break) that undergoes a sudden increase in velocity dispersion, such as that experienced during the onset of planet migration \citep{Fraser2009-divot}. 
\textcolor{black}{When the velocity dispersion rises, objects suddenly have enough kinetic energy to disrupt larger bodies than those they could shatter previously. Due to the transition from a steep to shallow distribution at the break, objects larger than the break see a relative increase in disruption rate than do objects smaller than the break, with the largest relative increase occurring for bodies of size just small enough to be disrupted by objects equal to the break radius. The result is preferential destruction of bodies $\sim10\times$ larger than the initial break radius.}
While the presence of a divot is not statistically required to match the observed ALF, its presence is compatible with observations, and, notably, is compatible with expectations of the formative and dynamical history of TNOs \citep{Fraser2009-divot}.

For completeness, we point out that the cold population appears to exhibit a different SFD than do the dynamically excited populations. 
\citet{Fraser2014} found that the largest cold classical objects have $D\sim400$~km and fall on the steep part of the SFD, with a slope $q\sim8$\textcolor{black}{\footnote{In terms of absolute magnitude, the brightest cold classical TNO is 79360 Sila-Nunam (1997 CS29), which has $H = 5.29$. Sila-Nunam is a roughly equal-mass binary in which each component has $D \approx 250$~km \citep{Grundy2012}.}}. Contrast this to the hot population SFD: the cold-classical SFD is missing the foot, and is much steeper brightward of the knee. 
This implies a significant difference in formation histories between those TNOs that appear to have formed in-situ, and those that were scattered outwards during Neptune's migration. The highest fidelity measure of the cold-classical SFD \citep{Kavelaars2021} demonstrates that this population is not well described by a broken power-law as in Eq.~(1), but rather is better fit by a modified exponential function, $N(<H)=10^{\alpha_SI(H-Ho)}\times \exp^{10^{\beta_SI(H-Ho)}}$, similar to the SFD produced in some simulations of formation via the streaming instability \citep{Li2019}. 
For the hot populations, it is likely that future observations will clarify whether power laws are reasonable descriptions of the SFD of the hot population. 

We also point out a tension in the populations of small, excited TNOs inferred from the ALF and the Pluto/Charon cratering record with that inferred from three serendipitous stellar occultations. Taken at face value, the occultations imply the existence of a population of small, $D\sim1$~km excited bodies that is more than an order of magnitude larger than implied from the former techniques \citep{Schlichting2012,Arimatsu2019,Parker2021}. A solution to this tension may be the presence of a very large population of small TNOs beyond $\approx$~50~au \citep{Shannon2021preprint}. It may also be the case that there is a population of small objects with extremely low albedos that have caused them to avoid optical detection. It may also be that some occultation events are not true detections, but rather are the result of unknown instrumental artifacts.

Centaurs and Jupiter-family comets provide another opportunity to probe the small-end size distribution of their parent population in the Kuiper Belt.
However, there are challenges to interpreting observations of both Centaurs and JFCs.
\textcolor{black} {The first arises from the lack of dedicated, well-characterized surveys, particularly for Centaurs. This is primarily due to the Centaurs' small numbers, a result of their short dynamical lifetimes on planet-crossing orbits, and their wide range of ecliptic latitudes owing to their dynamically hot inclination distribution. Centaurs are thus much less dense on the sky than TNOs. Most observational surveys are focused on detecting either the close Near Earth Asteroid population or the more distant TNO populations, so the biases in the observed Centaur population are typically not well understood (see, e.g., discussion in \citealt{Peixinho2020}).}
Another complication is that JFCs and some Centaurs are active.
For such bodies,  their observed magnitudes contain contributions from (and often are dominated by) the coma, rather than just the nucleus. 
Astronomers have tried to minimize coma signal  by taking observations when the comets are at aphelion and thus less active or inactive; however, there is always the possibility of unresolved coma that nonetheless affects the photometry \citep{Hui2018}.

With these limitations in mind, does the TNO size distribution match the JFC and Centaur size distributions? \citet{Lamy2004} review data on 65 ecliptic comets with effective diameters between 0.4 and 30 km, and infer a cumulative size distribution for $D \gtrsim 3$~km with $N(>D) \propto D^{-\gamma}$, where $\gamma = 1.9 \pm 0.3$, corresponding to $q \approx$2.6--3.2 for the differential size distribution and $\alpha \approx$ 0.3--0.4 for the luminosity function. \citet{Lamy2004} attribute the shallower size distribution seen for comets with $D \lesssim$~3 km to observational incompleteness and mass loss due to activity. \citet{Meech2004}, on the other hand, find that the flatter slope for small comets is not explained by observational bias. 

\citet{Snodgrass2011} performed Monte Carlo simulations of the JFC size distribution, accounting for uncertainties in photometry and the albedo, phase function, and shape of the nucleus. They inferred $\gamma = 1.92 \pm 0.20$ for nuclei with $D > 2.5$~km, consistent with \citet{Lamy2004}.

The largest survey to date of Jupiter-family comets was carried out by \citet{Fernandez2013}, who observed 89 JFCs with Spitzer and included nine other JFCs from the literature. 
They infer $\gamma \approx 1.9 \pm 0.2$ for $D > 3$~km, but note that, surprisingly, JFCs with $D > 6$~km and $q < 2$~au are still being found. \citet{Fernandez2013} agree with \citet{Meech2004} that the rarity of small comets is real.\footnote{Since the year 2000, the number of known Near-Earth Asteroids (defined as inactive bodies with $q < 1.3$~au) has increased by a factor of $\approx 30$, while the number of Near-Earth Comets has less than doubled. These numbers are tabulated by the Center for Near Earth Object Studies at {\tt https://cneos.jpl.nasa.gov/stats/totals.html}. Although biases in the discovery of asteroids and comets differ, the different discovery rates support the idea that small JFCs are intrinsically rare.} \citet{Jewitt2021} explains the scarcity of small short-period comets as the result of sublimation torques, which he estimates can spin up comets with perihelion distances between 1 and 2~au to rotational disruption in $25 (D/1 \, {\rm km})^2$~years, i.e., only a few orbits. 
The inferred SFD slopes for JFCs agree with those commonly adopted for the excited TNO parent populations, but the SFD of the Jupiter Trojans is more commonly used as a proxy for the TNO SFD  for comet-sized bodies\textcolor{black}{, as more Trojans are known than JFCs, and Trojans are inactive } \citep{Jewitt2000, Yoshida2019, Yoshida2020}.

Using data from the Deep Ecliptic Survey, \citet{Adams2014} found $\alpha = 0.42 \pm 0.02$ (i.e., $\gamma = 2.1 \pm 0.1$) for seven Centaurs with absolute magnitudes between 7.5 and 11, corresponding to 30~km $\lesssim D \lesssim 170$~km for an assumed albedo of 0.06. OSSOS discovered 15 Centaurs with $a < 30$~au, $q > 7.5$~au, and $H_r < 13.7$ (i.e., $D > 10$~km for an albedo of 0.06). \citet{Nesvorny2019-ossos} found that his planetary instability models predicted $11 \pm 4$ such Centaurs if he assumed $\gamma = 2.1$, similar to the slope inferred for small Trojans \citep{Wong2015}. 
In general, the debiased SFDs of the JFCs and Centaurs are consistent with those measured for the TNO populations that feed the JFCs, but a detailed comparison cannot yet be made. We discuss this in Section~\ref{sec:future}.

\section{Outstanding Questions and Future Prospects}
\label{sec:future}

In this section, we highlight three outstanding questions regarding TNOs and their link to cometary populations. We point out the anticipated contributions of certain new and upcoming telescope facilities that should provide significant leaps forward in our understanding of these questions. This list should in no way be considered impartial or complete, but rather reflects problems that the authors find particularly pertinent.

The first outstanding problem we wish to highlight regards the compositions of TNOs, and by extension the Centaurs and JFCs that are fed from the Kuiper Belt. As expressed in Section~\ref{sec:kbo_ices}, despite more than 25 years of observations by the astronomical community, very little is known about the surfaces or internal compositions of TNOs. We are still ignorant of the nature of the reddening agent responsible for the colors of these icy bodies. Moreover, beyond a couple of exceptions, no detection has been made of signatures of the silicate materials that must be present on and inside TNOs. 

\begin{figure}[!ht]
    \centering
    \includegraphics[width=0.49\textwidth]{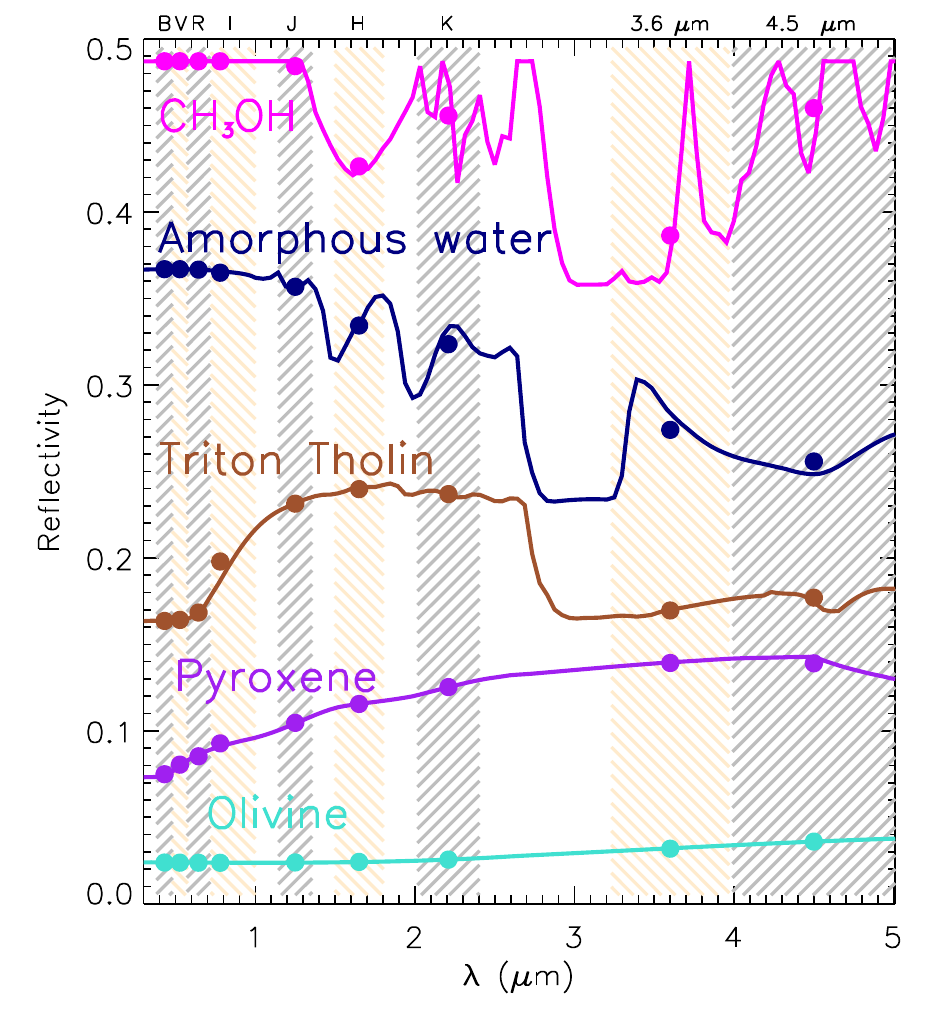}
    \caption{Spectra of various laboratory materials through the optical and NIR spectral range, reproduced from \citet{Fernandez-Valenzuela2021}. Photometric bands of the Johnson-Cousins system which are accessible from ground-based facilities, as well as the Spitzer-IRAC 3.6 and $4.5 \mbox{ $\mu$m}$ bands, are shown.}
    \label{fig:materials}
\end{figure}

The main reason for this sad state of affairs is the lack of identifying absorption features in the optical and NIR atmospheric transmission regions. It seems that whatever materials are present on the surfaces of TNOs, they mutually mask any features that fall in this wavelength range. See, for example, the spectral model of \textcolor{black}{the Centaur Pholus} in Figure~\ref{fig:TNOSpectra}, where tholin and olivine mask the strong absorption features each would exhibit on their own.

Longer wavelengths hold the potential to reveal some strong absorption features. For example, many of the ice species known or suspected to exist on TNOs exhibit deep and broad absorption features in the $2.5\lesssim \lambda \lesssim 4.5 \mbox{ $\mu$m}$ range (see Figure~\ref{fig:materials}).  This spectral region is very difficult to observe for these faint bodies with current technology, especially ground-based because of contamination from the Earth's atmosphere. It is for this reason that JWST holds the massive potential to revolutionize our knowledge of the compositions of TNOs. 

\begin{figure}[htp]
    \centering
    \includegraphics[width=0.49\textwidth]{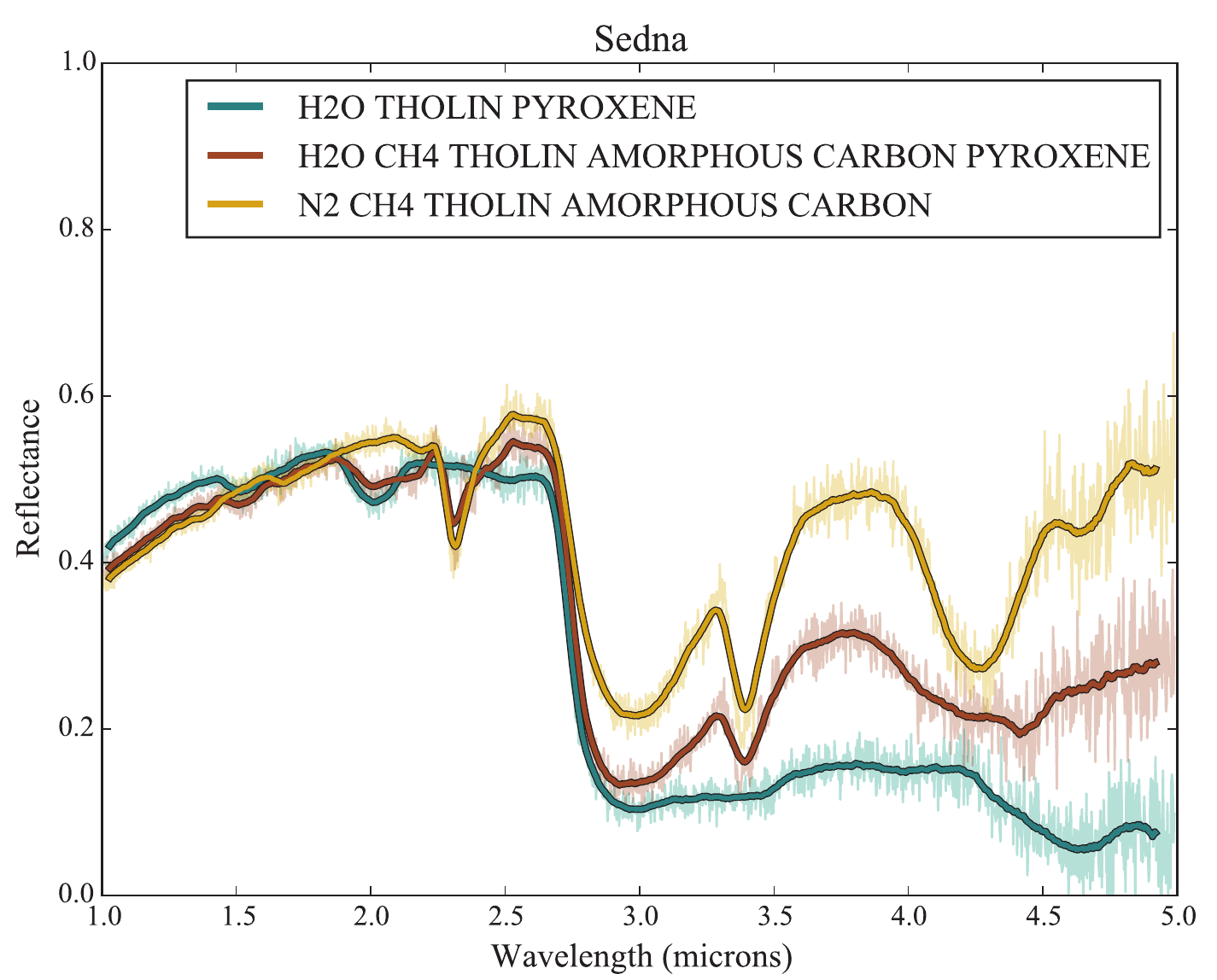}
    \caption{Spectral models of the TNO (90377) Sedna, using the silicate pyroxene, the organic material Titan tholin, CH$_4$, and amorphous carbon, and N$_2$ and H$_2$O ices. Reproduced from \citet{Parker2016}. \label{fig:sedna_model}}
\end{figure}

Most insight into the compositions of TNOs will come from spectral observations with the JWST-NIRSpec instrument, which will provide unprecedented sensitivity across a critical wavelength range out to 
5 \mbox{$\mu$m}. In Figure~\ref{fig:sedna_model} we reproduce a figure from \citet{Parker2016}, which presents three different spectral models, all of which are broadly compatible with the spectrum of the \textcolor{black}{distant dwarf planet} (90377) Sedna. Each model is wildly different in their compositional makeup, but are nearly indistinguishable in reflectance spectra at wavelengths $\lambda \lesssim 2.5 \mbox{ $\mu$m}$. At longer wavelengths, spectra from NIRSpec will be particularly useful in diagnosing a surface as organic-rich or organic-poor, as 
the C-H and C-N vibrational fundamental and overtone bands fall in this range \citep[see][for recent discussions]{Roush2004,Izawa2014}. Icy species, including H$_2$O, CH$_3$OH, CH$_4$, N$_2$, CO, and CO$_2$, also exhibit absorption features in this range that should be readily apparent in high-quality NIRSpec observations \citep[see Figure~\ref{fig:materials} of this chapter, and Figure~4 of ][]{Fernandez-Valenzuela2021}.

Longer wavelength spectra in the mid-infrared range $5<\lambda<28 \mbox{ $\mu$m}$  will come from the JWST-MIRI instrument, where fine-grained silicate materials exhibit emission features \citep{Martin2022}. While objects in the Kuiper Belt are too distant, and therefore too faint, for observations with MIRI, many Centaurs should be bright enough. \textcolor{black}{Spitzer observed 20 Centaurs in the mid-infrared \citep{Lisse2020}, but obtained spectral data over the range 7.5--38~$\mu$m for }only one, (8405) Asbolus \citep{Barucci2008-ssbn}. That spectrum, \textcolor{black}{which suggests the presence of fine-grained silicates on the Centaur's surface, is broadly similar to the spectra of three Jupiter Trojans observed by Spitzer \citep{Dotto2008}. }

\textcolor{black}{Unfortunately, the low signal-to-noise of Spitzer's spectrum of Asbolus makes it impossible to infer the nature of its putative silicates. }
The MIRI spectra should be of sufficient quality to not only definitively identify silicate emissions on many Centaurs, but also to identify specific properties of the silicates, such as whether those silicates are amorphous or crystalline in nature.

\begin{figure}[htpb]
    \centering
    \includegraphics[width=0.49\textwidth]{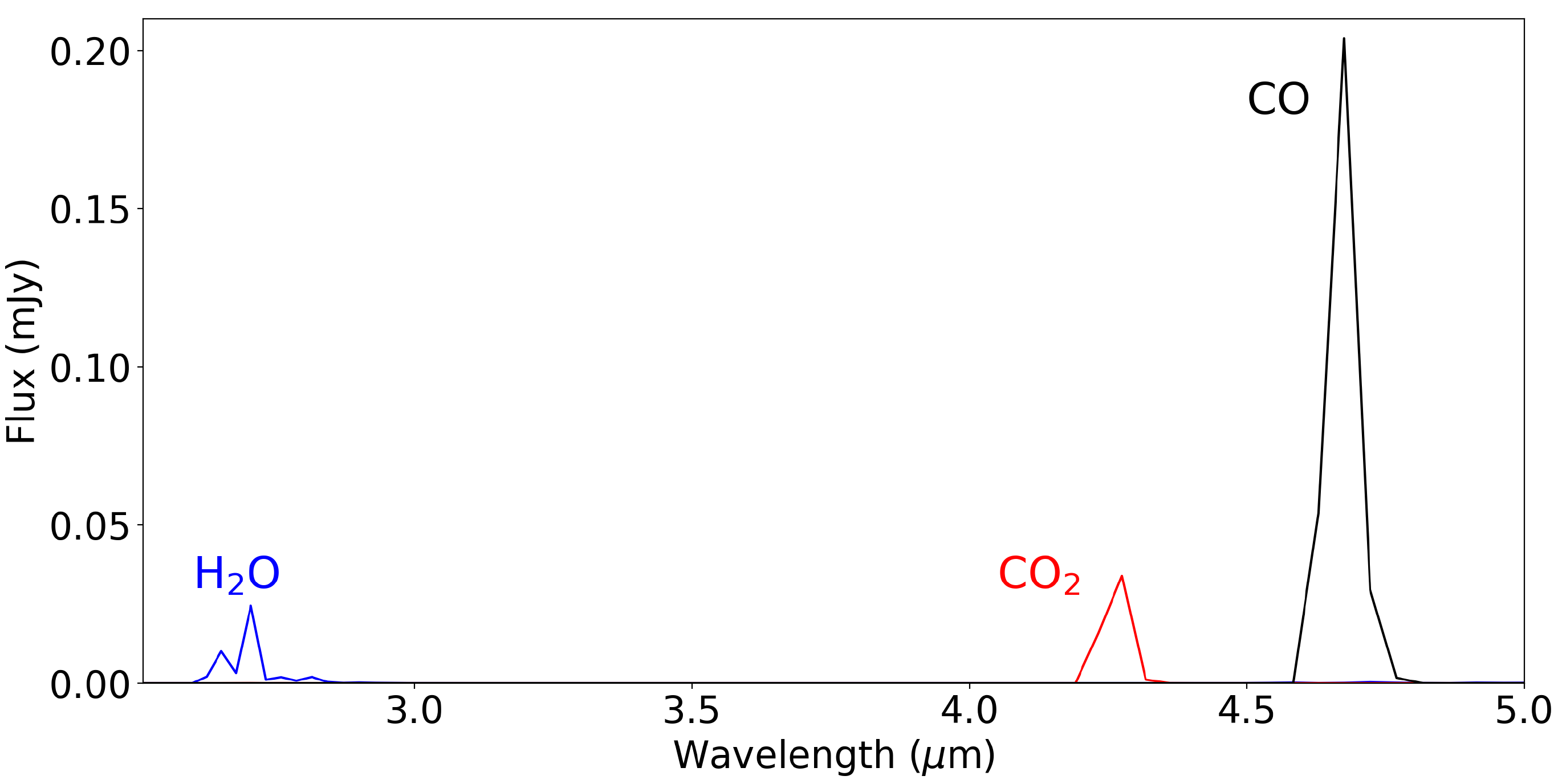}
    \caption{Predicted JWST NIRSpec model gas emissions for an active Centaur with a similar coma compositional profile as 29P  (A. McKay, personal communication). The model was calculated using the NASA Planetary Spectrum Generator ({\tt https://psg.gsfc.nasa.gov/} \citet{Villanueva2018}) assuming AKARI production rates for CO and H$_2$O from \citep{Ootsubo2012} and inferred Q(CO$_2$) from Harrington Pinto et al. (under review).} 
    \label{fig:spectrumgenerator}
\end{figure}

With regard to the volatile component, the JWST NIRSpec-IFU spectrograph is likely to either detect CO$_2$ emission in active Centaurs, or set strong upper limits, possibly up to two orders of magnitude less \textcolor{black}{than the limit of $3.5 \times 10^{26}$ for 29P achieved by }AKARI \citep{Ootsubo2012} [A. McKay, personal communication]. In addition, CO and H$_2$O emission will also be reachable with the instrument (see Figure \ref{fig:spectrumgenerator}). As discussed in Section~\ref{sec:centaur-activity}, CO$_2$ is reduced with respect to CO in the comae of objects beyond 3 au, a trend that increases with heliocentric distance (Harrington Pinto et al., under review). Thus, if other Centaurs are like 29P, then their comae may be also remarkably depleted in CO$_2$. JWST will be capable of detecting the 1.5, 2.0 and 3.0 $\mu$m bands of water, with the 3-$\mu$m band being the one most likely to be detected in the comae of active Centaurs, and the 2-$\mu$m band being the next likely.

An important, unanswered question about Centaur gaseous comae is whether comparing the CO, CO$_2$, and H$_2$O mixing ratios in cometary and Centaur comae is indeed a good match to compositional models of nuclei, or instead is more influenced by outgassing behavior at different heliocentric distances, due to differences in sublimation of the two species. This is a significant opportunity for the modeling community. 

The second outstanding problem we wish to highlight is the discrepancy between the SFD of the JFCs and that predicted from our knowledge of the source populations feeding the JFCs (see Sections~\ref{sec:origins_tnos} and ~\ref{sec:cent-to-JFCs}). While the slopes of the SFDs between the two populations seem to align (Section~\ref{sec:sfds}), there still appears disagreement between the availability of the source populations, and the observed density of JFCs.
This problem was highlighted by \cite{Volk2008}, who found that TNOs following the \citet{Bernstein2004, Bernstein2006} SFD for faint TNOs could produce only $\approx$~1\% of the observed JFCs.
While modern models of the supply of JFCs from the TNOs match the orbital distributions of the comets well \citep[e.g.][]{DiSisto2009, Brasser2013, Nesvorny2017}, the problem of directly, quantitatively linking the TNO and JFC population remains frustrated by several factors: the imprecise measurement of the SFDs of the scattered disk objects and other TNO populations, especially down to comet-sized objects; the unknown ratio of active to inactive JFCs, i.e., the fading problem discussed in Section~\ref{sec:sfds}; and the incomplete knowledge of the orbital structure of the most distant TNO populations and the interplay between the scattering and detached populations (discussed in Section~\ref{sec:tno_sources}). 

The Vera C. Rubin Telescope's Legacy Survey of Space and Time (LSST) will provide many key insights towards resolving this problem. 
One of the major advances from LSST will be a much improved census of the solar system's small bodies, down to a brightness threshold of $r\sim24.5$, with well-understood observational biases.
LSST will increase the inventory of known TNOs with well-determined orbits by at least an order of magnitude, detect large numbers of comet-sized ($d\lesssim10$~km) Centaurs, and increase the observed JFC population \citep{LSST2009}.
These observations will yield important new constraints on the detailed orbital distributions of all three populations.
LSST will also provide 10 years of monitoring for each discovered JFC and Centaur, on a 4--8~day cadence. 
This monitoring will be fundamental in detecting the onset and turnoff of activity as comets move between aphelion and perihelion, and measuring the vigor of the activity. 
It will also yield improved insights into the beginning stages of activity in the Centaur region.
Importantly, the monitoring will enable a robust derivation of the SFD of JFCs when they are inactive, which is what is required to make a robust comparison with the Centaur and TNO populations.    

A final issue is the possible presence of bodies more massive than Pluto and Eris (0.002--0.003$M_\oplus$) in the distant Kuiper Belt. Even with its sublunar mass, Pluto can destabilize bodies in the 3:2 MMR \citep{Nesvorny2000}. \citet{Gladman2006} investigated how an Earth-mass body, ten Mars-mass bodies (each $\approx 0.1 M_\oplus$), or both would raise the inclinations and perihelion distances of some objects in the primordial Kuiper Belt and produce detached bodies like Sedna. \citet{Lykawka2008} proposed that a planet with mass $\approx 0.5 M_\oplus$ excited the early Kuiper Belt and evolved into a stable orbit with $a > 100$~au, $q > 80$~au, and an inclination between 20 and 40$^\circ$. \citet{Trujillo2014} discovered a TNO, 2012~VP$_{113}$, with $a \approx 266$~au and a perihelion distance even larger than Sedna's, $q = 80$~au. They noted that the twelve TNOs with $a > 150$~au and $q > 30$~au appeared to cluster in argument of perihelion, and proposed that a 2--15M$_\oplus$ body could produce the clustering. This idea is now known as the ``Planet Nine'' hypothesis \citep{Batygin2019, Brown2021} and invokes a 5–-10$M_\oplus$ body with a perihelion distance $\approx 300$~au to shape the orbits of TNOs with $a > 250$~au. Such a planet would even affect the orbital distributions of JFCs and HTCs \citep{Nesvorny2017}. Well-characterized TNO surveys \citep{Shankman2017, Kavelaars2020, Napier2021} find that the orbital distribution of ``extreme'' TNOs does not require the existence of Planet Nine, but neither can it (or less massive perturbers) be ruled out. Assuming an albedo between 0.20 and 0.75, \citet{Brown2021} predict that Planet Nine's most likely R magnitude is $\approx 20$, but it could be as faint as magnitude 25. \textcolor{black}{Searches in the far-IR with archival IRAS and Akari data have also been carried out recently \citep{Rowan-Robinson2022, Sedgwick2022}.} Observations in the next decade with the Vera Rubin Observatory and other large telescopes \textcolor{black}{such as Subaru, the VLT, and Gemini} should clarify the orbital distribution of distant TNOs and may finally detect the long-speculated trans-Neptunian planet.

\vskip .5in
\noindent \textbf{Acknowledgments} \\

KV acknowledges support from NSF (grant AST-1824869) and NASA (grants 80NSSC19K0785, \\
80NSSC21K0376 and 80NSSC22K0512). LD thanks the Cassini Data Analysis Program for support.
DN would like to acknowledge support from the Emerging Worlds program.
This material is based in part on work done by MW while serving at the National Science Foundation.
We'd especially like to thank Rosita Kokotanekova and Alan Fitzsimmons for their verbal contributions to this chapter. We thank Adam McKay and Gareth Williams for useful discussions. We also thank Sunao Hasewaga for providing data tables. 

\bibliographystyle{sss-three.bst}
\bibliography{kbos-to-jfcs.bib, dont-edit-refs-from-ads.bib}

\end{document}